\pdfoutput=1
\documentclass[usenatbib, twocolumn]{mnras}
\usepackage[dvipdfmx]{graphicx}
\usepackage[dvipdfmx]{color}
\usepackage{amsmath,amssymb}
\usepackage{dcolumn}
\usepackage{bm}
\usepackage{xcolor}

\newcommand{\beq}{\begin{equation}}
\newcommand{\beqa}{\begin{eqnarray}}
\newcommand{\eeq}{\end{equation}}
\newcommand{\eeqa}{\end{eqnarray}}

\newcommand{\simgt}{\lower.5ex\hbox{$\; \buildrel > \over \sim \;$}}
\newcommand{\simlt}{\lower.5ex\hbox{$\; \buildrel < \over \sim \;$}}

\newcommand{\bd}[1]{\mbox{\boldmath $#1$}}

\newcommand{\comment}[1]{{\color{blue}\bf{(#1)}}}
\newcommand{\change}[1]{{\color{black} #1}}

\usepackage{etoolbox}
\makeatletter
\patchcmd\@combinedblfloats{\box\@outputbox}{\unvbox\@outputbox}{}{%
   \errmessage{\noexpand\@combinedblfloats could not be patched}%
}%
\makeatother

\title[Baryonic effects on galaxy cluster mass profiles]
{Modelling Baryonic Effects on Galaxy Cluster Mass Profiles}

\author[M. Shirasaki, E.\ T.\ Lau, D.\ Nagai]
{Masato Shirasaki$^{1}$\thanks{E-mail: masato.shirasaki@nao.ac.jp},
Erwin T.\ Lau$^{2,3}$,
and
Daisuke Nagai$^{2,3}$
\\
$^{1}$Division of Theoretical Astronomy, National Astronomical Observatory of Japan, 
Mitaka, Tokyo 181-8588, Japan \\
$^{2}$Department of Physics, Yale University, 
New Haven, CT 06520, USA \\
$^{3}$Yale Center for Astronomy and Astrophysics, Yale University,
New Haven, CT 06520,USA \\
}
\begin{document}

\date{}

\pagerange{\pageref{firstpage}--\pageref{lastpage}} \pubyear{2017}

  \maketitle

\label{firstpage}

\begin{abstract}
Gravitational lensing is a powerful probe of the mass distribution of galaxy clusters and cosmology. However, accurate measurements of the cluster mass profiles \change{are} limited by 
\change{uncertainties in cluster astrophysics}.
%the still poorly understood cluster astrophysics. 
In this work, we present a physically motivated model of baryonic effects on the cluster mass profiles, which self-consistently takes into account the impact of baryons on the concentration as well as mass accretion histories of galaxy clusters. We calibrate this model using the {\em Omega500} hydrodynamical cosmological simulations of galaxy clusters with varying baryonic physics. Our model will enable us to simultaneously constrain cluster mass, concentration, and cosmological parameters using stacked weak lensing measurements from upcoming optical cluster surveys.

\end{abstract}


\section{INTRODUCTION}

Tracing the evolution of abundance of galaxy clusters over cosmic time
is a promising approach to constrain cosmological parameters, such as the amplitude of mass density fluctuations, the equation of state of dark energy, and possible extensions to the standard $\Lambda$CDM model
\citep[e.g., see][for review]{2011ARA&A..49..409A}.
The current cosmological constraints derived from cluster abundance
hinge on the accuracy of mass estimation and the systematic uncertainty due to inaccurate mass estimates already dominates the statistical error in current cluster-based cosmological constraints 
\citep[e.g.][]{
2009ApJ...692.1060V, 
%2010ApJ...722.1180V, 
%2011ApJ...732...44S, 
%2013ApJ...763..147B, 
2013JCAP...07..008H, 
%2014A&A...571A..20P, 
2014MNRAS.440.2077M, 
2016A&A...594A..24P, 
2016ApJ...832...95D}.

Gravitational lensing is a powerful technique for accurately measuring the cluster masses, because the lensing effect is purely induced by gravity, independent from the physical nature or the dynamical state of the gravitating mass.  Thanks to deep galaxy imaging surveys, it has become feasible to 
measure the mass density distribution around individual galaxy clusters
through gravitational lensing measurements \citep[e.g.][]{2008ApJ...685L...9B, 2008JCAP...08..006M, 2010PASJ...62..811O, 2012MNRAS.420.3213O, 2014ApJ...795..163U}. 

However, accurate cluster mass estimation requires controlling various systematic uncertainties in modelling the mass distribution of galaxy clusters. Baryonic processes, such as radiative cooling, star formation, and feedback from supernovae and active galactic nuclei, can modify the mass profiles of galaxy clusters \citep[e.g.,][]{2008ApJ...672...19R, 2010MNRAS.405.2161D,2014MNRAS.442.2641V,2015MNRAS.451.1247S, 2015MNRAS.452..343S}. The change in the mass distribution due to baryonic effects can be characterized by the change in the mass concentration of the cluster halo. This effect has been modelled using simple adiabatic contraction in response to the baryonic dissipation in the cluster center \citep[e.g.,][]{2011arXiv1108.5736G, 2012MNRAS.424.1244F}. However, these adiabatic contraction models did not take into the account the dependence of the cluster mass profiles (via halo mass concentration) on the mass assembly histories (MAH). 

The mass concentration of a dark matter (DM) halo has been shown to be correlated with its MAH \citep[e.g.,][]{2002ApJ...568...52W}. Recent work by \citet{2013MNRAS.432.1103L} showed that statistically the mass profile of a given DM halo is directly related to its MAH, through the inside-out halo growth: inner regions of the halo are formed earlier, while the outer regions are governed by late-time accretion. Disentangling the effects of MAH and baryonic effects on halo mass concentration are important for accurate weak lensing mass estimates. 

In this work, we extend the relation between mass concentration and MAH of cluster-size halo of \citet{2013MNRAS.432.1103L}, by incorporating the changes in mass concentration due to baryonic physics. We calibrate this model using the {\em Omega500} hydrodynamical cosmological simulations that include various baryonic physical processes. We then apply our model to stacked weak lensing measurements to constrain the effects of baryons on the mass concentration and weak lensing mass estimate. Our model provides a physically motivated framework for simultaneously constraining both baryonic physics and cluster masses with optical cluster surveys, such as the upcoming Large Synoptic Survey Telescope (LSST) survey. 

Our paper is organized as follows.
Section~\ref{sec:model} summarizes a theoretical model of the mass distribution and concentration of DM halo and its relation to the MAH, including the effects of baryons. A brief description of our hydrodynamical simulations is given in Section~\ref{sec:sim}.
In Section~\ref{sec:res}, we present the relation of \change{the} mass profile with MAH in the presence of baryonic processes and a detailed comparison with our model and simulations, and investigate baryonic effects on the halo concentration
for future weak lensing measurements of galaxy clusters. 
Our main results are summarized in Section~\ref{sec:conclusions}.

Throughout this paper, $\log$ represents the logarithm with base 10, while $\ln$ is the natural logarithm.

%---------------------------------------------------%

\section{Theoretical model} \label{sec:model}

%---------------------------------------------------%

%-------------------------------%
\subsection{Mass profile}
%-------------------------------%

To characterize the mass distribution of gravitationally bounded objects, we assume averaged mass density profile $\rho(r)$ is spherically symmetric and is only a function of the cluster-centric radius $r$. 
%This profile can also be expressed in terms of the enclosed mean density $\langle \rho \rangle$,
%\beqa
%\langle \rho \rangle (r) &=& \frac{M(<r)}{(4\pi/3)r^3}, \label{eq:enc_rho_def} \\
%M(<r) &=& \int_{0}^{r}\, {\rm d}x\, 4\pi x^{2}\rho(x),
%\eeqa
%where $M(<r)$ is the enclosed mass within radius of $r$. 
%Using Eq.~(\ref{eq:enc_rho_def}), 
%we can express the enclosed mass as function of the enclosed mean density $\langle \rho \rangle$, denoted as $M(\langle \rho \rangle)$.
Numerical simulations have shown that spherically averaged density profiles of DM halos are well approximated by scaling a simple formula proposed in \citet{Navarro1996}.
The profile is referred as the NFW profile and it is given by
\beqa
\frac{\rho_{\rm NFW}(r)}{\rho_{\rm crit}}=\frac{\delta_{c}}{(r/r_s)(1+r/r_s)^2},
\label{eq:def_NFW}
\eeqa
where 
$\rho_{\rm crit}(z)$
is the critical density of the universe at redshift of $z$, 
$r_s$ is a scale radius, 
and
$\delta_c$ is a dimensionless characteristic density of the halo. 
These parameters $\delta_c$ and $r_s$ can be condensed into one parameter, 
the concentration $c_{\Delta}=r_{\Delta}/r_s$, through the definitions of the halo mass 
\beqa
M_{\Delta} &\equiv& \frac{4\pi}{3} \Delta \, \rho_{\rm crit}(z) r^3_{\Delta}, \label{eq:SOmass_def} \\
M_{\Delta} &=& \int_{0}^{r_{\Delta}}\, {\rm d}x\, 4\pi x^{2}\rho_{\rm NFW}(x), \label{eq:SOmass_NFW}
\eeqa
where $\Delta$ is the over-density parameter, and 
Eqs~(\ref{eq:SOmass_def}) and (\ref{eq:SOmass_NFW}) reduce to
\beqa
\delta_{c} = \frac{\Delta}{3}\frac{c^3_{\Delta}}{\left[\ln (1+c_{\Delta})-c_{\Delta}/(1+c_{\Delta})\right]}.
\eeqa
\change{In the following, we simply write the density concentration parameter as $c$ and 
use $c_{\Delta}$ as necessary.}

\change{For the NFW density profile with a given $\Delta$, we have the expression of the mass 
profile, denoted as $M(\langle \rho \rangle)$, in terms of the enclosed mean density, $\langle \rho \rangle$,}
%for the NFW density profile 
by solving the following equations;
\beqa
\frac{M(x)}{M_{\Delta}} &=& \frac{Y(c x)}{Y(c)},
\,\,\,\,\,\,\,\,\,\,
\frac{\langle \rho \rangle(x)}{\Delta \rho_{\rm crit}} 
= \frac{1}{x^3}\frac{Y(cx)}{Y(c)}, \label{eq:NFW_scaled}
\eeqa
where 
$x=r/r_{\Delta}$ and $Y(u)=\ln(1+u)-u/(1+u)$.
%Eq.~(\ref{eq:NFW_scaled}) shows that
%the normalizations of $M/M_{\Delta $ and 
%$\langle \rho \rangle / \Delta \rho_{\rm crit}$
%enable us to find the best expression of the mass profile $M(\langle \rho \rangle)$ 
%in terms of the single parameter $c_{\Delta}$.

%-------------------------------%
\subsection{Relation between mass profile and accretion history}
%-------------------------------%

Recently, \citet{2013MNRAS.432.1103L} 
showed that there exists a one-to-one relation 
between mass profile $M(\langle \rho \rangle)$ and the mass accretion history (MAH) of the halo.
The MAH is commonly defined as
\change{the spherical over-density mass}
of the main progenitor as a function of redshift, 
denoted as $M_{\rm prog}(z)$.
%The average MAH can be expressed by a NFW-like expression similar to Eq.~(\ref{eq:NFW_scaled}):
\change{Analogously to Eq.~(\ref{eq:NFW_scaled}), we assume the average MAH can be given by the following
parametric expression:}
\beqa
\frac{M_{\rm prog}(z)}{M_{\rm prog}(z_o)}
= \frac{Y(c_{\rm MAH}q)}{Y(c_{\rm MAH})},
\,\,\,\,\,\,\,\,\,\,
\frac{\rho_{\rm crit}(z)}{\rho_{\rm crit}(z_o)}
= \frac{1}{q^3}\frac{Y(c_{\rm MAH}q)}{Y(c_{\rm MAH})}, \label{eq:MAH_scaled}
\eeqa
where 
\change{$q$ is a parameter of this model} and $z_o$ is the redshift for galaxy clusters of interest.
The assumption of
Eq.~(\ref{eq:MAH_scaled}) works for DM halos 
with various masses and redshifts on average 
\change{in dark-matter-only simulations}
\citep{2013MNRAS.432.1103L}.
In addition, the ``concentration'' parameter $c_{\rm MAH}$ is found to be 
tightly correlated with the mass concentration of $c$.
This tight correlation between $c_{\rm MAH}$ and $c$ can be summarized as 
\beqa
\langle \rho \rangle (r_s) \propto \rho_{\rm crit}(z_{s}), \label{eq:CDM_prof_MAH_relation}
\eeqa
where a scaled redshift $z_{s}$ is given by
\beqa
M_{\rm prog}(z_s) = M(<r_s). \label{eq:scaled_redshift}
\eeqa

%-------------------------------%
\subsection{Modelling baryonic effects on mass concentration}
%-------------------------------%

To model the effects of baryons on the mass concentration, we generalize the relation of Eq.~(\ref{eq:CDM_prof_MAH_relation}) as 
\beqa
\log\left(\frac{\langle\rho\rangle (r_s)}{\rho_{o}}\right)
= \alpha_{0} + \alpha_{1} \log\left(\frac{\rho_{\rm crit} (z_s)}{\rho_{o}}\right),
\label{eq:general_mass_prof_MAH_relation}
\eeqa
where $\rho_{o}$ is the critical density of the universe at $z=z_{o}$.
Throughout this paper, we refer to
Eq.~(\ref{eq:general_mass_prof_MAH_relation}) as the mass-profile-MAH relation.

\change{
Substituting Eqs~(\ref{eq:NFW_scaled}) and (\ref{eq:MAH_scaled}) into Eq.~(\ref{eq:general_mass_prof_MAH_relation}),
one can find a unique relation 
between $c$ and $c_{\rm MAH}$, and it is independent of redshift $z_o$. 
According to \citet{2013MNRAS.432.1103L}, 
the relation can be approximately expressed as
\beqa
c = a_1 \left( 1+a_{2} c_{\rm MAH}\right)^{a_3},
\label{eq:cs_relation}
\eeqa
where $a_{1}, a_{2}$ and $a_3$ are constants, which can be evaluated for given $\alpha_0$ and $\alpha_1$.
}

\change{
In this paper, we assume that baryonic effects will change $\alpha_0$ and $\alpha_1$ in Eq.~(\ref{eq:general_mass_prof_MAH_relation}), but $c_{\rm MAH}$ is
determined primarily by gravitational structure formation alone and insensitive to baryonic effects.
In other words, the impact of baryonic effects on mass concentration can be controlled
by three parameters $a_{1}, a_{2}$ and $a_{3}$ in Eq.~(\ref{eq:cs_relation}),
while $c_{\rm MAH}$ can be responsible for mass-redshift-cosmology dependence.
We then follow the procedure developed in \citet{2014MNRAS.441..378L} to compute 
the mass concentration $c$ for arbitrary masses, redshifts, cosmologies given Eq.~(\ref{eq:cs_relation})\footnote{
We also summarize the procedure to compute 
the mass concentration $c$ in Appendix~\ref{app:mah_model}.}.
}

In the following section, we will examine these assumptions with high-resolution hydrodynamical cosmological
simulations with varying baryonic physics.

\section{SIMULATIONS}\label{sec:sim}

%---------------------------------------------------%

%-------------------------------%
\subsection{The {\em Omega500} simulation}
%-------------------------------%
In this work, we analyze the mass-limited sample of galaxy clusters extracted from the {\it Omega500} simulation series. 
The simulations assume 
flat $\Lambda$CDM model with the WMAP five-year results \citep{2009ApJS..180..330K}:
$\Omega_{\rm m0} = 0.27$ (matter density), $\Omega_{\rm b0} = 0.0469$ (baryon density),
\change{$H_{0} \equiv 100h = 70\, {\rm km}\, {\rm s}^{-1}{\rm Mpc}^{-1}$} (Hubble constant),
and $\sigma_{8} = 0.82$ (the mass variance within a sphere with a radius of 8 $h^{-1}\, {\rm Mpc}$).
The simulation is performed using the Adaptive Refinement Tree (ART)
$N$-body+gas-dynamics code 
\citep{1999PhDT........25K, 2002ApJ...571..563K, 2008ApJ...672...19R}, 
which is an Eulerian code that uses adaptive refinement in space and time and non-adaptive refinement in mass to achieve the dynamic range necessary to resolve the cores of halos formed 
in self-consistent cosmological simulations.  
The simulation volume has a comoving box length 
of 500~$h^{-1}\, {\rm Mpc}$, resolved using a uniform $512^3$ root grid and 8 levels of mesh refinement, implying a maximum comoving spatial resolution of 3.8~$h^{-1}\, {\rm kpc}$.

The {\it Omega500} simulation series consists of three runs under different baryonic physics;
non-radiative (NR) hydrodynamics as in \citet{2014ApJ...782..107N},
the run with additional baryonic physics, such as radiative cooling, star formation (CSF), 
and further including feedback from active galactic nuclei (AGN).
In the NR run, we have treated the ICM as a non-radiative gas and ignored additional baryonic physics.
The CSF run includes metallicity-dependent radiative cooling,
star formation, thermal supernova feedback, metal enrichment and advection, 
which are based on the same subgrid physics modules in 
\citet{2007ApJ...668....1N}, to which we refer the reader for more detail. 
The AGN run includes the CSF physics, mentioned above, 
plus a subgrid thermal AGN feedback module, similar to the one adopted in in \citet{booth_etal09}. 
First, supermassive blackholes (SMBH) are seeded as particles with an initial mass of $10^5h^{-1}M_\odot$ 
at the centres of DM halos with $M_{500c} > 2 \times 10^{11}h^{-1}M_\odot$. 
These SMBH grow via mergers and gas accretion with a rate given by a modified Bondi accretion model and return the feedback energy as a fraction of the accreted rest mass energy ($\epsilon = 0.2$) into the environment in the form of thermal energy.  
We also impose a minimum heating temperature, 
$T_{\rm min} = 10^7$~K, requiring that the SMBH store enough feedback energy 
until they accumulate enough energy to heat neighbouring gas cells, each by an amount of $T_{\rm min}$ to keep the injected thermal feedback energy from radiating away. 

Cluster-sized halos are identified in the simulation 
using a spherical overdensity halo 
finder described in \citet{2014ApJ...782..107N}. 
We define the three-dimensional mass of cluster-size halos using the spherical overdensity criterion with $\Delta=500$ in Eq.~(\ref{eq:SOmass_def}). We select DM halos with 
$M_{\rm 500c}\geq 3 \times 10^{14}\, h^{-1}M_{\odot}$ at 
$z=0$ and re-simulate the box with higher resolution DM particles 
in regions of the selected halos with the ``zoom-in'' technique \citep{2001ApJ...554..903K}, resulting in an effective mass resolution of $2048^{3}$, corresponding to a DM 
particle mass of  $1.09\times10^{9}\, h^{-1} M_{\odot}$, inside a spherical region with the cluster-centric radius of three times the virial radius for each halo. 
After the selection with $M_{\rm 500c}\geq 3 \times 10^{14}\, h^{-1}M_{\odot}$, we found 63, 80, 82 halos in the NR, CSF, and AGN runs, respectively.

%%%%%%%%%%%%%%%%%%%%%%%%%%%%%
\begin{figure}
\centering
\includegraphics[width=0.80\columnwidth, bb=0 0 511 550]
{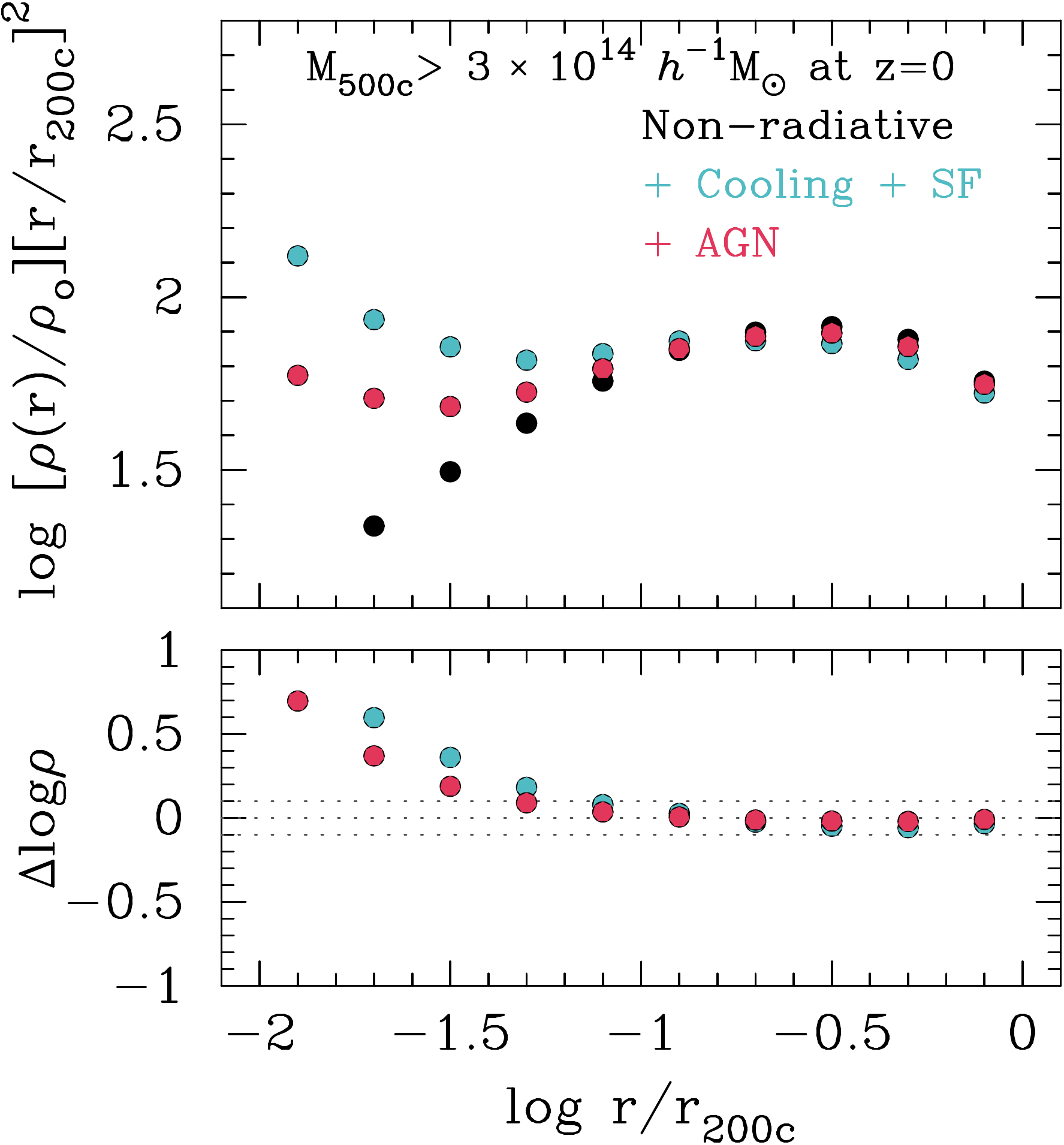}
\caption{
	The baryonic effect on mass density profile of 
	cluster-sized DM halos.
	The colored points show the average mass profile 
	for the mass-selected sample with 
	$M_{\rm 500c}\ge3\times10^{14}\, h^{-1}M_{\odot}$ 
	at $z=0$.
	Black, cyan and red points represent the results from the NR, CSF, and AGN runs.
	Residuals from the NR case are shown in the bottom panel.
	The dashed lines at the bottom highlight the level of $\pm0.1$ dex.
	}
\label{fig:profiles_sim}
\end{figure} 
%%%%%%%%%%%%%%%%%%%%%%%%%%%%%

Note that our CSF simulation suffers from the well-known 
``overcooling'' problem, where the simulation
over-predicts the amount of central stellar mass by a factor of $\sim 2$. As such, the results of our NR and CSF runs can be used to {\em bracket} uncertainties associated with baryonic effects.
The effects of baryonic physics, such as radiative gas cooling, 
star formation and energy feedback from supernovae and AGN are important in the cluster core regions.

Figure~\ref{fig:profiles_sim} highlights the impact of baryonic physics on the mass density profile around cluster-sized halos.
In the figure, we show the mass density profile averaged over 
the mass-selected sample in three {\it Omega500} runs.
We found the additional radiative effects can increase the mass density profile in the region of $r/r_{\rm 200c}<0.1$.
We also confirmed these contractions in the mass profile are 
consistent with the modified adiabatic contraction model
as in \citet{2011arXiv1108.5736G} with appropriate parameters
(see Appendix~\ref{app:mass_contraction_model} for details).

%-------------------------------%
\subsection{Fitting Density Profiles and Mass Accretion Histories}
%-------------------------------%

%-----------------------------------------------------------%
\subsubsection{Mass density profile as a function of radius}
%-----------------------------------------------------------%

In this section, we examine the relation of the enclosed mass profile $M(\langle \rho \rangle)$ and MAH \citep{2013MNRAS.432.1103L} in the presence of baryonic physics using the {\em Omega500} hydrodynamic cosmological simulations. 

We first obtain the best representation of
spherically averaged mass density profile $\rho(r)$ and then 
set a scale radius $r_s$ to define the relation as in Eq.~(\ref{eq:CDM_prof_MAH_relation}). 
Specifically, we compute the spherically averaged density profiles 
of cluster-sized halos in the {\em Omega500} simulations.
We then construct radial profile by $\log$-spaced binning 
in the range of $10 <r /(h^{-1} {\rm kpc})<10^4$ with $99$ bins.
To find the best representation of $\rho(r)$,
we fit using the non-linear 
least-squares Levenberg-Marquardt algorithm \citep{1992nrfa.book.....P}
by minimizing the $\chi^{2}$-fitting metric defined as 
\beqa
\chi^{2} = \sum_{i=1}^{N_{\rm bin}}
\left[\ln\rho(r_{i})-\ln\rho_{\rm NFW}(r_{i}|\, r_s, \, \delta_c)\right]^2,
\eeqa
where $N_{\rm bin}$ is the number of radial bins.
When performing $\chi^2$ fitting, we consider the radial range of 
$0.15<r/r_{200c}<1$, avoiding the core region in order to mitigate uncertainties associated 
with physical modelling of cluster core regions (e.g., see Figure~\ref{fig:profiles_sim}).

After finding the best-fitted parameter of $r_s$ for each halo,
we derive the scaled redshift $z_s$ from the information of MAH
and using Eq.~(\ref{eq:scaled_redshift}).

%-------------------------------%
\subsubsection{Mass profile as a function of density}
%-------------------------------%
Next, we consider the enclosed mass profile as a function of density.
For the matter density profile around cluster-sized halos at $z=0$, 
we introduce the expression of the enclosed mass profile as a function of the enclosed mass density, denoted as $M(\langle \rho \rangle)$.
Assuming the NFW profile, we fit the profile $M(\langle \rho \rangle)$
with single parameter $c_{\Delta}$ once we properly normalize
$M$ and $\langle \rho \rangle$ as shown in Eqs~(\ref{eq:NFW_scaled}).
To find the best representation of $M(\langle \rho \rangle)$,
we use a $\chi^{2}$-fitting metric defined as
\beqa
\chi^{2} = \sum_{i=1}^{N_{\rm bin}}
\left[\ln m(\hat{\rho}_{i})-\ln m_{\rm NFW}(\hat{\rho}_{i}|\, c_{\Delta})\right]^2,
\eeqa
where $m=M/M_{\Delta}$, $\hat{\rho}=\langle \rho \rangle/(\Delta \rho_{\rm crit}(z=0))$,
and $m_{\rm NFW}(\hat{\rho})$ 
is given by the solution of the parametric form given by 
Eq.~(\ref{eq:NFW_scaled}). 
\change{We use the logarithmic binning in the range of $1\le\hat{\rho}\le10^2$ 
with the bin size of $\Delta \log \hat{\rho} = 0.2$.}
When performing $\chi^2$ fitting on the mass profile, 
we adopt $\Delta=200$ and limit the density range of 
\change{$1 <\langle \rho \rangle/[200 \, \rho_{\rm crit}(z=0)] < 35$}, 
where the latter condition corresponds to the excision of cluster core regions with $r/r_{200c}\simlt0.15$.

%-------------------------------%
\subsubsection{Mass Accretion History}
%-------------------------------%
Finally, we measure MAH of individual halos as a function of redshift. Here, we define the cluster mass by $M_{\rm 500c}$. Using Eq.~(\ref{eq:MAH_scaled}), we perform a $\chi^2$ fitting to the MAH by minimizing the following metric as
\beqa
\chi^{2} = \sum_{i=1}^{N_{\rm bin}}
\left[\ln m_{\rm prog}(\hat{\rho}_{{\rm prog}, i})
-\ln m_{\rm prog, NFW}(\hat{\rho}_{{\rm prog}, i}|\, c_{\rm MAH})\right]^2,
\eeqa
where 
$m_{\rm prog} = M_{\rm prog}(z)/M_0$,
$\hat{\rho}_{\rm prog} = \langle \rho \rangle/\rho_{\rm crit}(z=0)$, and
$m_{\rm prog, NFW}(\hat{\rho}_{\rm prog})$ 
is given by the solution of Eq.~(\ref{eq:MAH_scaled}).
\change{We use the logarithmic binning in the range of $1\le\hat{\rho}_{\rm prog}\le10$ 
with the bin size of $\Delta \log \hat{\rho}_{\rm prog}= 0.2$.}

%---------------------------------------------------%

\section{Fitting the Mass-profile-MAH model}\label{sec:res}

%---------------------------------------------------%

%---------------------------------------------------%
\subsection{Impacts of baryonic physics on the Mass-profile-MAH relation}
\label{subsec:mass_prof_MAH}
%---------------------------------------------------%

%%%%%%%%%%%%%%%%%%%%%%%%%%%%%
\begin{figure}
\centering
\includegraphics[width=0.80\columnwidth, bb=0 0 519 471]{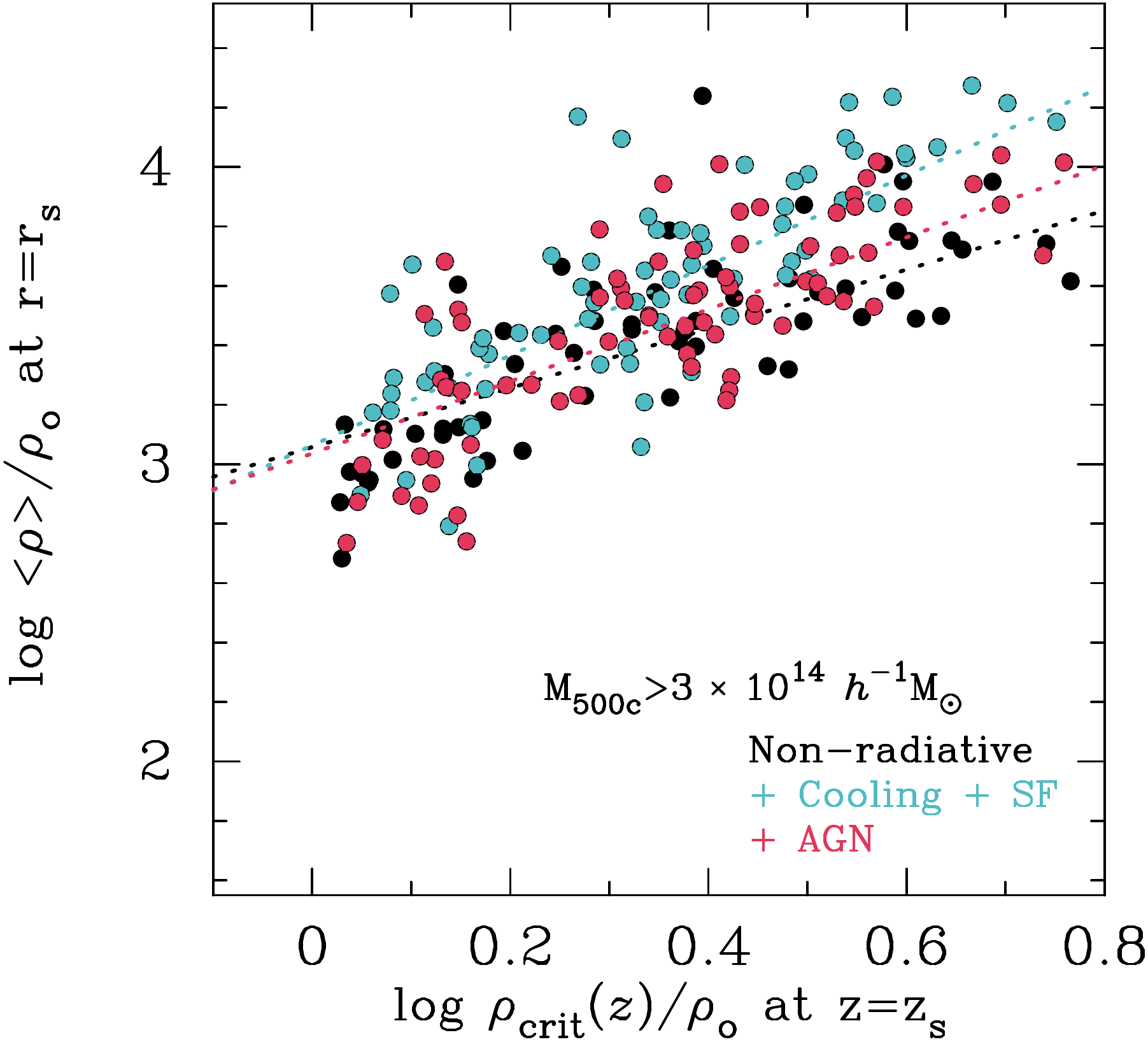}
\caption{
	Relation between mass profiles at $z=0$ 
	and accretion histories for mass-limited haloes 
	($M_{\rm 500c}\ge3\times10^{14}\, h^{-1}M_{\odot}$)
	in our simulations with various baryonic physics.
    \change{
    The vertical axis shows the mean enclosed densities within the scale radius, 
    $r_{s}$, of the NFW profile. 
    The horizontal axis is for the critical density of the universe at the time 
	when the main progenitor's mass equals the present mass enclosed within $r_s$.
	Each point represents densities normalized 
	by the critical density in the universe at $z=0$, denoted as $\rho_{o}$.
    }
	%The scaled density of mass profile is defined as the mean enclosed densities within the scale radius, 
    %$r_{s}$, of the NFW profile.
	%For MAH, we define the scaled density as
	%the critical density of the universe at the time 
	%when the main progenitor's mass equals the present mass enclosed within $r_s$.
	Black, cyan, and red points indicate individual halos from the NR, CSF, and AGN simulations, respectively, while the colored dashed lines are the best-fitted scaling relations for each run.
}
\label{fig:rho_MAH}
\end{figure} 
%%%%%%%%%%%%%%%%%%%%%%%%%%%%%

First, we examine the correlation between the enclosed mass density $\langle \rho \rangle$ and the critical density $\rho_{\rm crit}$ in MAH for individual halos. Figure~\ref{fig:rho_MAH} shows the scatter plot of $\langle \rho \rangle$ at $r=r_s$ and the critical density at $z=z_s$, 
where we define $z_s$ as in Eq.~(\ref{eq:scaled_redshift}).
In this figure, the different coloured points represent the results for three different physics runs (NR, CSF, AGN) in {\em Omega500}.
Performing a linear least-square fitting to 63 clusters in
our NR sample at $z=0$, the best-fit normalization and slope for the scaling relation between
$\log \langle \rho \rangle (r_s)$ and $\log \rho_{\rm crit}(z_s)$ (Eq.~\ref{eq:general_mass_prof_MAH_relation}) is
$(\alpha_0, \alpha_1) = (3.057\pm 0.289, 0.996\pm0.528)$, where the error values
indicate the $1\sigma$ errors. The best-fit slope for NR haloes is consistent with 
the results obtained using the dark-matter-only simulations by \citet{2013MNRAS.432.1103L}.

Once including baryonic effects associated with gas cooling and star formation in the CSF run, we find $\langle \rho \rangle (r_s)$ tends to become higher for a given $\rho_{\rm crit}$.
The best-fit parameters for our CSF sample is $(\alpha_0, \alpha_1) = (3.063\pm 0.354, 1.513\pm0.528)$, which indicates marginal $1\sigma$ deviation from the DM-only linear relation given by the Eq.~(\ref{eq:CDM_prof_MAH_relation}).
For the AGN run, the best-fit parameters are $(\alpha_0, \alpha_1) = (3.036\pm 0.320, 1.212\pm0.504)$.
Although the mass-profile-MAH relation for individual haloes
has a large scatter, as shown in Figure~\ref{fig:rho_MAH},
we find that the mean mass-profile-MAH relation is sensitive to baryonic physics, as noted by the significant change in slope of the mean relation, $\alpha_{1}$ in Eq~(\ref{eq:general_mass_prof_MAH_relation}).

We also quantify the intrinsic scatter of the $\langle \rho \rangle (r_s)-\rho_{\rm crit}(z_s)$ relation as
\beqa
\sigma^2_{\rm int} = \frac{1}{N-1}\sum_{i=1}^{N}
\left[\log \langle \rho \rangle_{i} - \log \langle \rho \rangle_{\rm fit}
\left(\rho_{\rm crit, i}\right)\right]^2,
\eeqa
where $N$ is the number of halos and $\langle \rho \rangle_{\rm fit}$
is the best-fit relation for each run. The best-fit parameters and scatters are summarized in Table~\ref{tab:alphas}.

%%%%%%%%%%%%%%%%%%%%%%%%%%%%%
\begin{figure*}
\centering
\includegraphics[width=1.5\columnwidth, bb=0 0 570 416]
{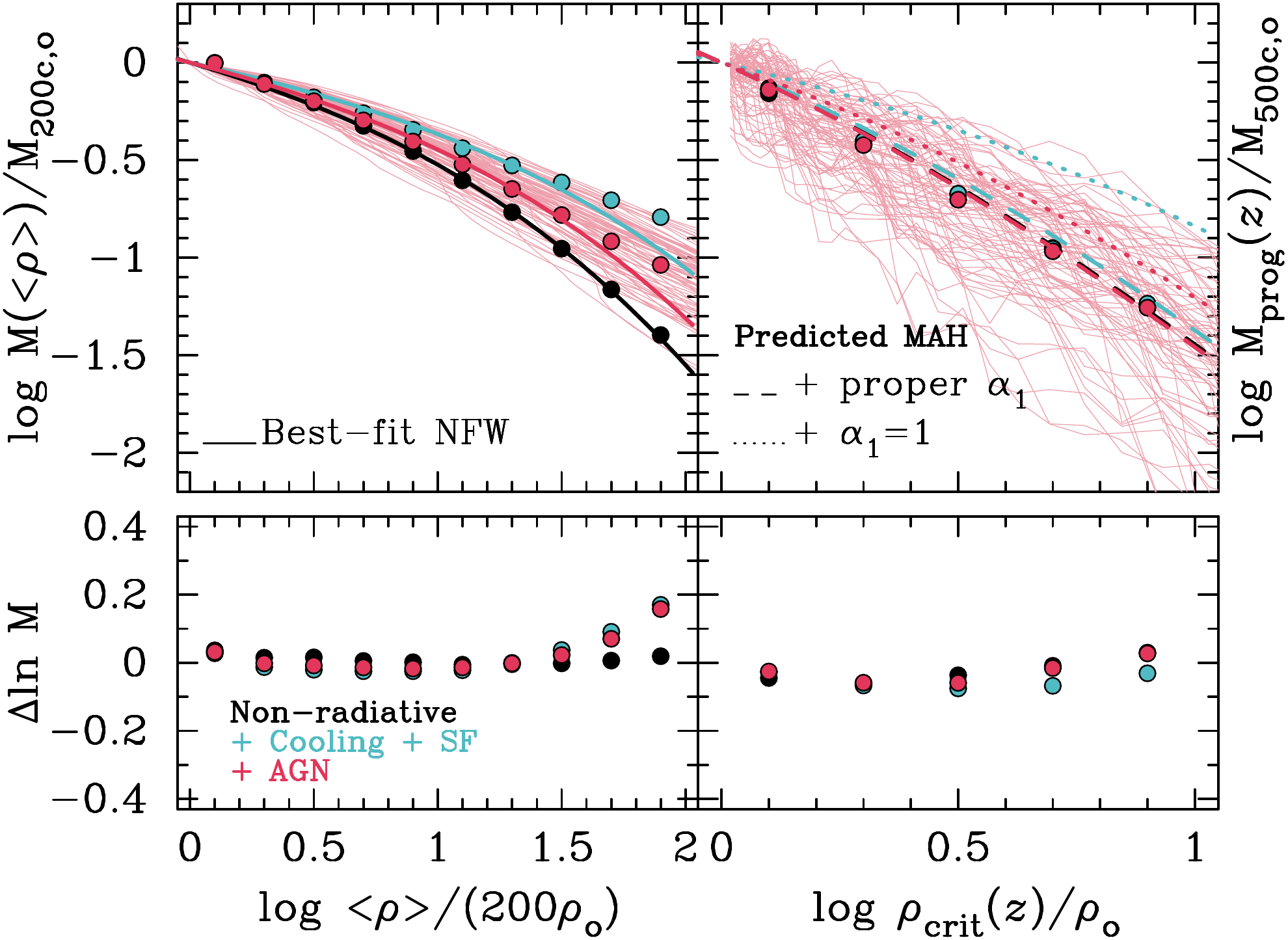}
\caption{
 	Mass profile as a function of mean enclosed density
	and accretion history as a function of the critical 
    density of the universe. 
    Top left panels show the mass profile, 
    while the MAH is presented in top right.
	We consider the mass-limited sample with 
    $M_{\rm 500c}\ge3\times10^{14}\, h^{-1}M_{\odot}$
	at $z=0$ for three different simulations.
    In each panel, black, cyan, and red point or line are the results of the NR, CSF, and AGN runs, respectively.
    In both top panels, the points show the average values 
    over the sample with different baryonic physics runs.
    Solid lines in the top left panel show 
    the NFW profiles fitted to the average mass profiles for each run,
	while the dashed lines in the right panel 
    represent MAH predicted
	by the relation between mass profiles and accretion histories as in
    Figure~\ref{fig:rho_MAH}.
    For comparison, the cyan and red dotted lines in the top right panel
    show the model predicted by 
    Eq~\ref{eq:general_mass_prof_MAH_relation} with $\alpha_1=1$.
	The bottom left panel shows the residual of the measured mass profile 
    from best-fit NFW profile.
	In the bottom right panel, we show the difference between 
    the measured and predicted MAH.
    Note that the thin red lines in both top panels show 
    the result for individual halos in the AGN run, illustrating
    the typical scatter in the mass profiles.
	}
\label{fig:mass_MAH_comp}
\end{figure*} 
%%%%%%%%%%%%%%%%%%%%%%%%%%%%%

%%%%%%%%%%%%%%%%%%%%%%%%%%%%%
\begin{table*}
\centering
\caption{
	Relation of concentration parameters 
	in mass profile and accretion history
	based on the mass-profile-MAH relations 
	$ (\alpha_{0},\alpha_1,a_1,a_2a_3)$
	and the intrinsic scatter around the best fit 
	$\sigma_{\rm int}$,
	measured in the {\it Omega500} simulations.
    Note that the parameters are valid for $\Delta=200$ in mass profile and when MAH is defined as the spherical over-density mass with $\Delta=500$.
	}
\begin{tabular}{@{}lcccccccl}
\hline
\hline
Run & $\alpha_{0}$ & $\alpha_1$ & $a_1$ & $a_2$ & $a_3$  & $\sigma_{\rm int}$ \\ \hline 
NR  & $3.057\pm 0.289$ & $0.995\pm 0.528$ & $2.840$ & $0.646$ & $1.010$  & $0.206$ \\ \hline 
CSF & $3.063\pm 0.354$ & $1.513\pm 0.528$ & $3.342$ & $0.836$ & $1.438$  & $0.217$ \\ \hline 
AGN & $3.036\pm 0.320$ & $1.212\pm 0.504$ & $2.864$ & $0.764$ & $1.190$ & $0.209$ \\ \hline 
\end{tabular}\label{tab:alphas}
\end{table*}
%%%%%%%%%%%%%%%%%%%%%%%%%%%%%

\change{
Since $\langle \rho \rangle/\rho_o \propto c^3 [\ln(1+c)-c/(1+c)]^{-1}$ for the NFW profile,
it is expected that there exists a scaling relation between $c$ and $\rho_{\rm crit}(z_s)$.
We confirmed this expectation with {\it Omega500} simulations.
The scaling relation can be approximated as
\beqa
\log c
= \beta_{0} + \beta_{1} \log\left(\frac{\rho_{\rm crit} (z_s)}{\rho_{o}}\right), \label{eq:c_MAH_relation}
\eeqa
where $(\beta_0, \beta_1) = (0.402\pm 0.147, 0.467\pm 0.279)$ for the NR run,
$(0.405\pm 0.164, 0.651\pm 0.243)$ for the CSF run, 
and $(0.384\pm 0.165, 0.572\pm 0.261)$ for the AGN run.
Here the error values indicate the $1\sigma$ errors and $c$ is defined by $r_{200c}/r_s$ at $z=0$.
The intrinsic scatter around Eq.~(\ref{eq:c_MAH_relation}) in base 10 logarithmic space
is found to be 0.106, 0.103, 0.107 for the NR, CSF, and AGN runs, respectively.
}

%---------------------------------------------------%
\subsection{Testing the mass-profile-MAH relation model with hydrodynamical simulations}
\label{sec:detailed_test_relation}
%---------------------------------------------------%

Next, we test the accuracy of our mass-profile-MAH relation model by checking whether the model can reproduce the average MAH from the mass profile. To do this, we first measure the average mass profile for our mass-limited sample of simulated clusters and fit the average mass profile $M(\langle \rho \rangle)$ with the NFW profile. We then infer the MAH by using this average mass profile and the mass-profile-MAH relation given in Eq.~(\ref{eq:general_mass_prof_MAH_relation}), and compare the result with the true MAH directly measured from the simulation. 

Figure~\ref{fig:mass_MAH_comp} shows the comparison of the average MAH (left panel) and the predicted MAH from the best-fit NFW profile (right panel) for the mass-limited sample with $M_{\rm 500c}\ge3\times10^{14}\, h^{-1}M_{\odot}$ in the {\it Omega500} runs with different baryonic physics. 
We find that the predicted MAH from mass profile is in reasonable agreement with the average MAH for all three runs. 
The accuracy of our model is at a level of 0.1 dex in the range of $1< \rho_{\rm crit}/\rho_{o} < 10$, corresponding to $0<z\simlt2.5$.

We also quantify the importance of baryonic physics in estimating the MAH from mass profile. In Figure~\ref{fig:mass_MAH_comp}, the differences between dashed and dotted lines in the top right highlight the importance of including the baryonic effects on the $\langle \rho \rangle (r_s)-\rho_{\rm crit}(z_s)$ relation.
If applying a simple relation of $\langle \rho \rangle (r_s) \propto \rho_{\rm crit}(z_s)$ to CSF/AGN runs, we find an estimate of the MAH at $\rho_{\rm crit}/\rho_{o}\simeq3$ or $z\sim1$ will be biased at the level of $\sim0.4$ and 0.2 dex for CSF and AGN runs, respectively. 

As observed in the three different physics runs in the {\em Omega500} simulation, baryonic effects can modify the concentration in mass density profile without affecting the overall MAH represented by $\rho_{\rm crit}(z_s)$. This can be understood by the relative importance of gravitational physics versus baryonic physics during different phases of the MAH of the halo. In the early phase of the MAH, where the cluster halo is rapidly accreting its mass, baryonic effects are expected to be sub-dominant to gravitational physics. This means that $\rho_{\rm crit}(z_s)$ in the mass-profile-MAH relation, which represents a typical density of the halo at the early formation phase, is relatively insensitive to the details of baryonic processes. At the later stage, where the halo is slowly accreting its mass, most of the accreted mass is deposited onto the outer regions of the halo. Baryonic physics dominates in the relatively quiescent cluster core region, rendering the average density within the scale radius, $\langle \rho \rangle (r_s)$, to become more
susceptible to the baryonic effects. 
As shown in Figure~\ref{fig:profiles_sim}, 
$\langle \rho \rangle (r_s)$ increases for runs with larger baryon dissipation. 
Therefore, baryonic effects increase $\langle \rho \rangle (r_s)$ without changing $\rho_{\rm crit}(z_s)$\footnote{
\change{Our simulations indicate that baryonic effects change $\langle \rho \rangle (r_s)$ and $r_s$ so that
Eq~(\ref{eq:scaled_redshift}) remains valid.}
}.

\subsection{Predicting halo concentration}
\label{subsec:From_MAH_to_prof}
%---------------------------------------------------%

\if0
%%%%%%%%%%%%%%%%%%%%%%%%%%%%%
\begin{figure}
\centering
\includegraphics[width=0.8\columnwidth, bb=0 0 507 550]
{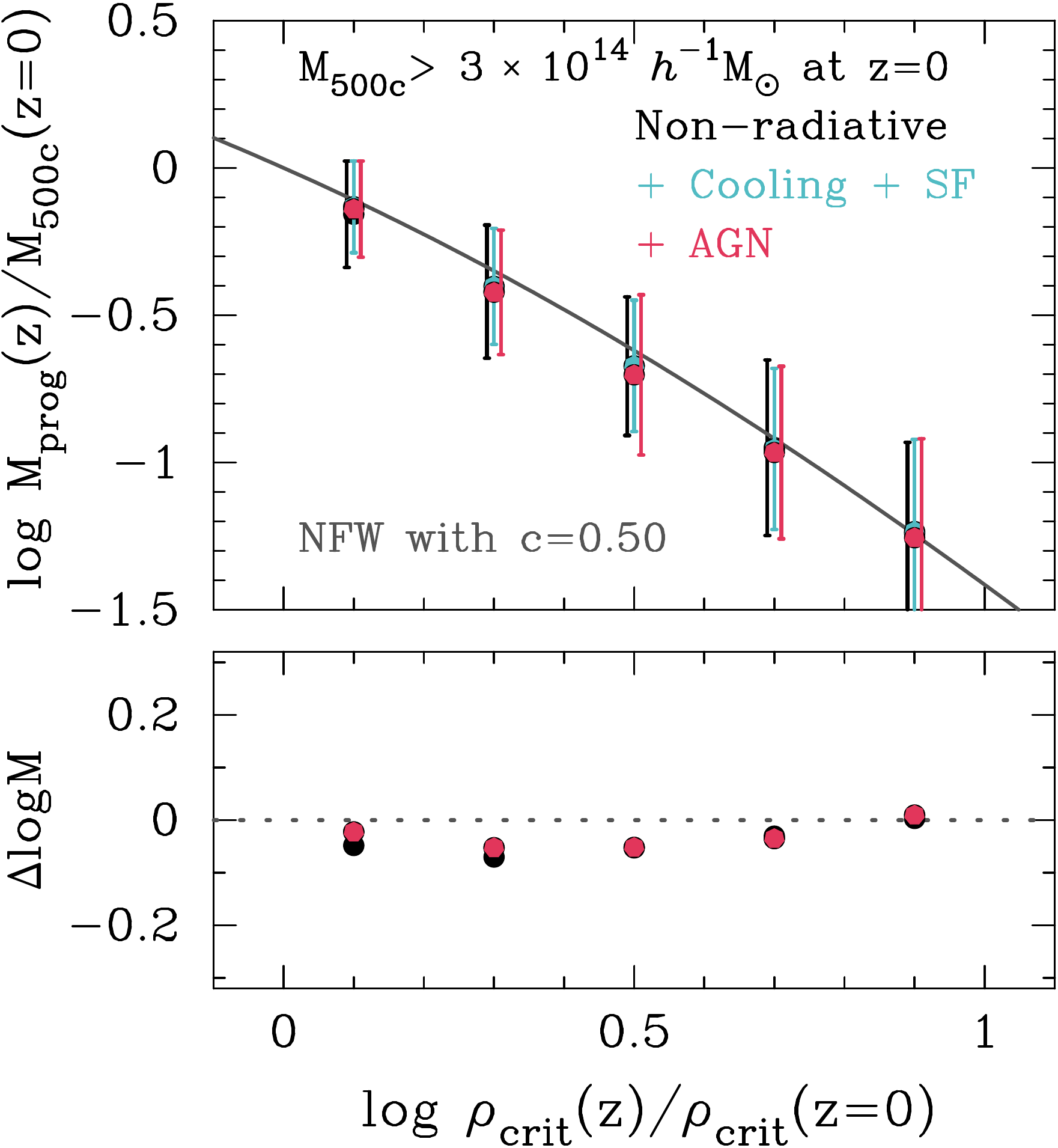}
\caption{
	The baryonic effect on MAH of 
	cluster-sized DM halos.
	The colored points show the average mass profile 
	for the mass-selected sample with 
	$M_{\rm 500c}\ge3\times10^{14}\, h^{-1}M_{\odot}$ 
	at $z=0$, whereas the error bars are for the standard deviations.
	Black, cyan, and red points represent the results of NR, CSF, and AGN runs, respectively.
	Gray line shows an NFW prediction with concentration of 0.5.
	Residuals from the NFW prediction are shown in the bottom panel.
	}
\label{fig:MAH_sims}
\end{figure} 
%%%%%%%%%%%%%%%%%%%%%%%%%%%%%
\fi

%%%%%%%%%%%%%%%%%%%%%%%%%%%%%
\begin{figure}
\centering
\includegraphics[width=0.8\columnwidth, bb=0 0 512 475]
{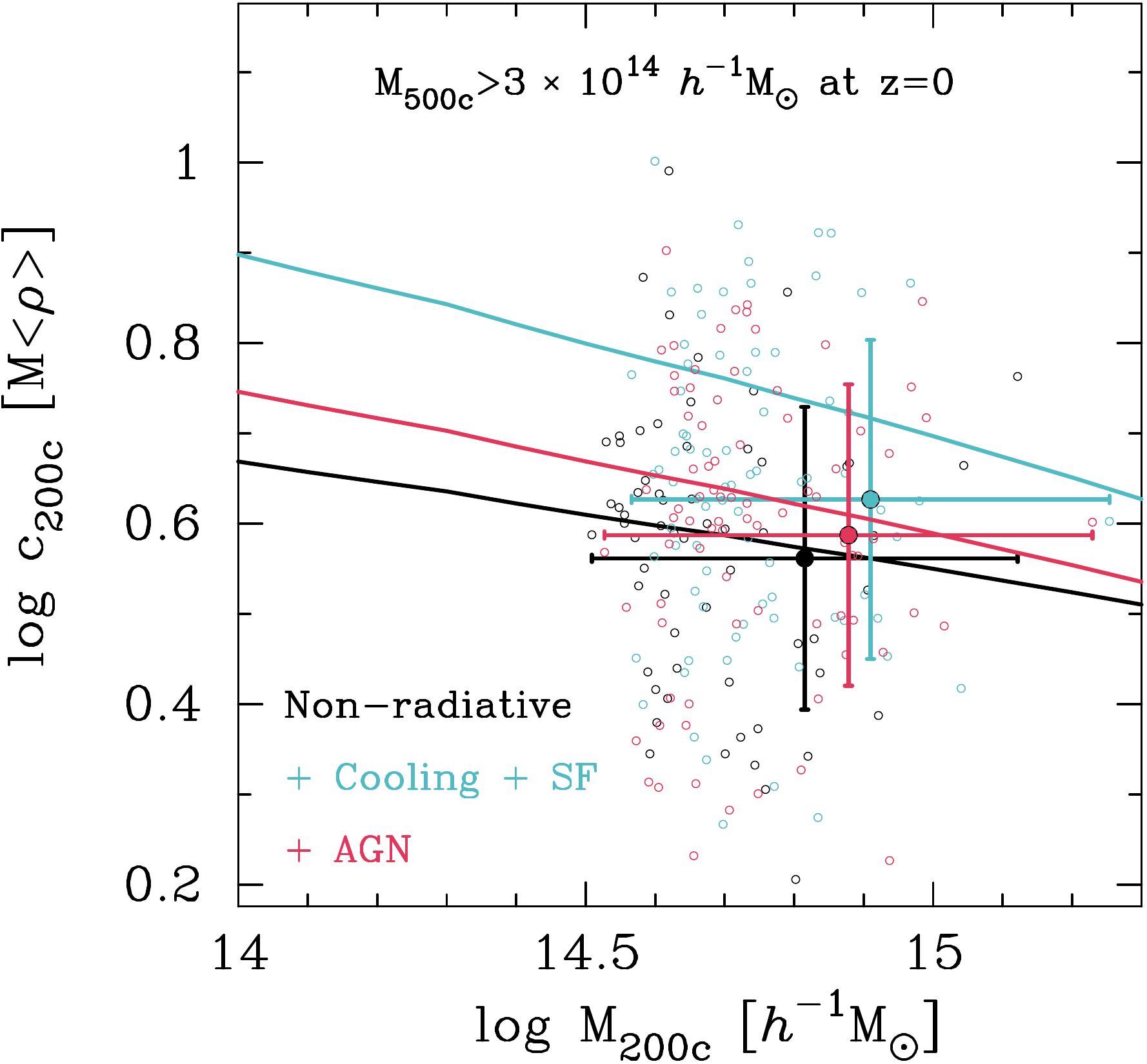}
\caption{
	Mass dependence in concentration parameters of mass profile. 
    The small, open, colored points are the best-fitted concentration parameters
	from NFW fits to mass profiles and the average concentrations 
	are shown in the large filled points.
	The different curves represent our model assuming
	the mass-profile-MAH relation and the universal MAH proposed by \citet{2002MNRAS.331...98V}.
	The baryonic effect on halo concentration in simulated mass profiles
	can be explained primarily by the difference in the mass-profile-MAH relations.
	}
\label{fig:cMAH_cprof_Mass}
\end{figure} 
%%%%%%%%%%%%%%%%%%%%%%%%%%%%%

%%%%%%%%%%%%%%%%%%%%%%%%%%%%%
\begin{figure}
\centering
\includegraphics[width=0.8\columnwidth, bb=0 0 509 475]
{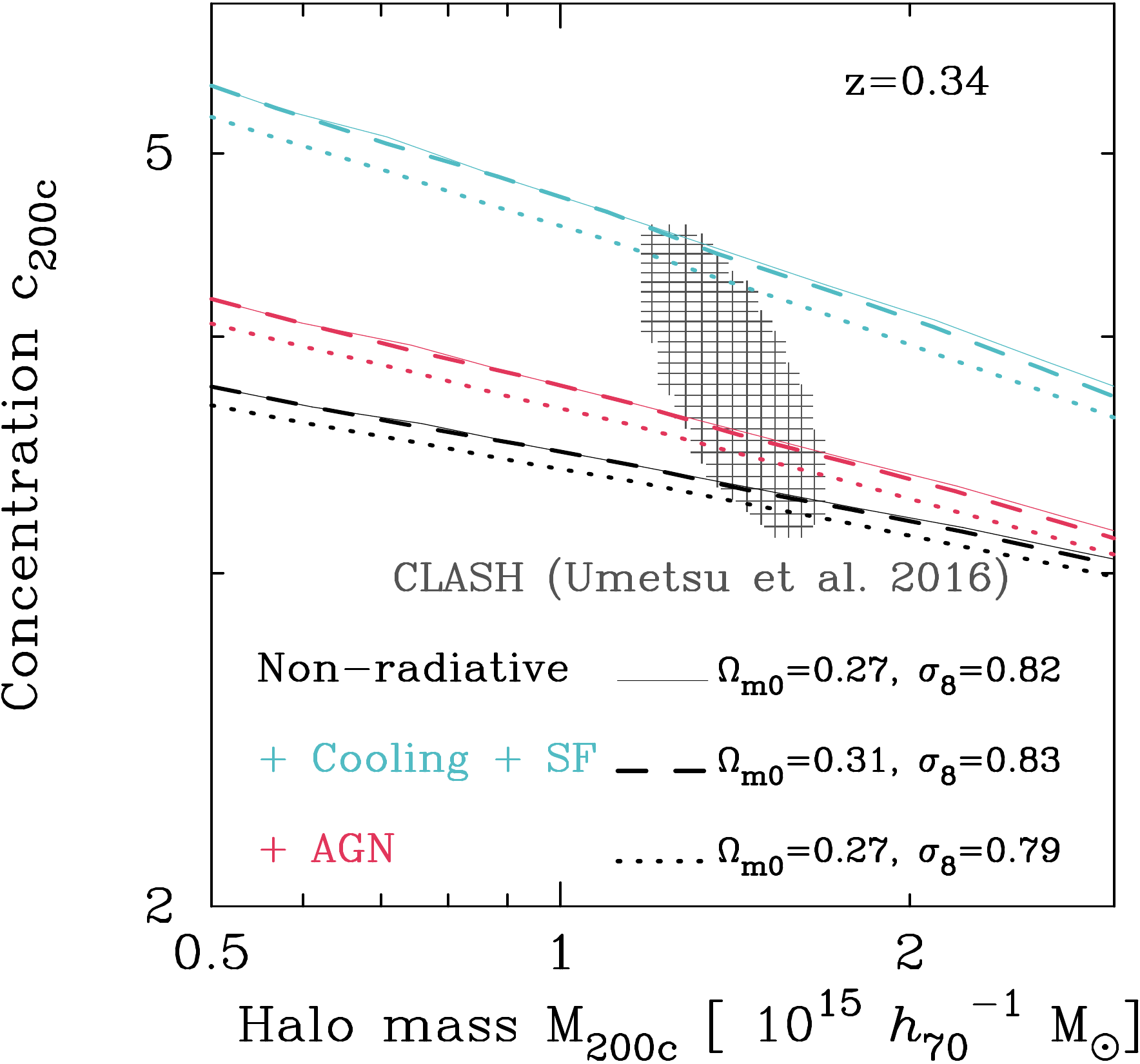}
\caption{
	Comparison of concentration of cluster-sized DM halos
	with our model and the observational constraint from 
	Hubble Space Telescope and Subaru Telescope
	\citep{2016ApJ...821..116U}.
	Different colored lines show the baryonic effect 
	on the halo concentrations, while the solid, dashed, and dotted lines
	reflect the dependence on cosmological parameters.
	The solid and dashed lines are the best-fit cosmological models 
    consistent with WMAP \citep{Hinshaw2013} and Planck \citep{2016A&A...594A..13P}, respectively.
	The dotted lines correspond to the cosmological model that
	is in agreement with 
	recent cosmic shear measurement \citep{2017MNRAS.465.1454H}. 
	The gray hatched region shows the $2\sigma$ constraints
	obtained from the combined strong and weak lensing analyses 
	of X-ray selected clusters
	\citep[see][for details]{2016ApJ...821..116U}. 
	We consider the redshift of 0.34 and $h=0.7 h_{70}$
	for the comparison. 
	}
\label{fig:cprof_Mass_comp_obs_model}
\end{figure} 
%%%%%%%%%%%%%%%%%%%%%%%%%%%%%

Armed with the well-tested mass-profile-MAH relation, we can predict the concentration of galaxy clusters by taking into account the effects of baryon physics on their total mass profiles. Figure~\ref{fig:cMAH_cprof_Mass} shows the comparison between $c_{200c}$ predicted by our model and those measured directly in simulations, demonstrating that our model can explain a trend of increasing halo concentration in the runs with enhanced baryon dissipation. Specifically, for the mass-limited sample with $M_{\rm 500c}\ge3\times10^{14}\, h^{-1}M_{\odot}$ at $z=0$, our model predicts the concentration parameter of 3.77, 5.52, and 4.21 for NR, CSF, and AGN runs, respectively.
These values are consistent with the halo concentrations measured directly by fitting NFW profiles to the $\it Omega500$ simulations with $1\sigma$ uncertainty as shown in Figure~\ref{fig:cMAH_cprof_Mass}.
Note also that our model assumes that the baryonic effects on the average MAH of cluster-sized DM halos are negligible, and we confirmed that this assumption is tenable in cluster-sized halos (see the right panel in Figure~\ref{fig:mass_MAH_comp}). The mass-profile-MAH relation in Section~\ref{subsec:mass_prof_MAH} plays a central role in explaining 
\change{the sensitivity of the concentrations found in hydrodynamical simulations to the baryonic physics implemented.}

Figure~\ref{fig:cprof_Mass_comp_obs_model}
summarizes the predictions of our theoretical model of 
$c_{\Delta}-M_{\Delta}$ relation at $z=0.34$.
Here, we consider three representative cosmological models
that are consistent with recent CMB measurements by WMAP
\citep{Hinshaw2013}
and Planck
\citep{2016A&A...594A..13P},
and cosmic shear measurement by Kilo-Degree Survey
\citep{2017MNRAS.465.1454H}. 
We consider a different set of $\Omega_{\rm m0}$
and $\sigma_{8}$ as indicated in the legend.
The different colored lines are the results based on the mass-profile-MAH
relations calibrated by a set of three simulations with different baryonic physics. 
We find that $\sim10\%$ difference of $\Omega_{\rm m0}$ is unimportant for determining the $c_{\Delta}-M_{\Delta}$ relation,
while $\sim4\%$ difference of $\sigma_8$ is more prominent.
Instead, the impact of baryonic physics on $c_{\Delta}$
is found to be critical under the concordance $\Lambda$CDM cosmologies.
Hence, the precise measurement of $c_{\Delta}-M_{\Delta}$ relation
will play an essential role \change{in discerning how baryons affect the properties of galaxy clusters}.
For comparison, the hatched gray region in the figure 
also shows recent observational constraint 
of the $c_{\Delta}-M_{\Delta}$ relation for X-ray selected high mass clusters
\citep{2016ApJ...821..116U}.
Although the current constraint is in agreement 
with our models that include three different baryonic physics,
Figure~\ref{fig:cprof_Mass_comp_obs_model} demonstrates
a possibility of distinguishing these models 
if the measurement of $c_{\Delta}$
at lower mass scales is performed.

%---------------------------------------------------%

\section{Constraining Baryonic Effects with Stacked Weak Lensing Measurements}
\label{sec:application}

%---------------------------------------------------%

Our theoretical framework for predicting the mass profiles of gravitationally bound objects can be applied to various cosmological analyses. In this section, we consider how future stacked lensing analysis of galaxy clusters can constrain the effects of baryons on the concentration and mass profiles of galaxy clusters. 

%%%%%%%%%%%%%%%%%%%%%%%%%%%%%
\begin{figure*}
\centering
\includegraphics[width=1.2\columnwidth, bb=0 0 519 501]
{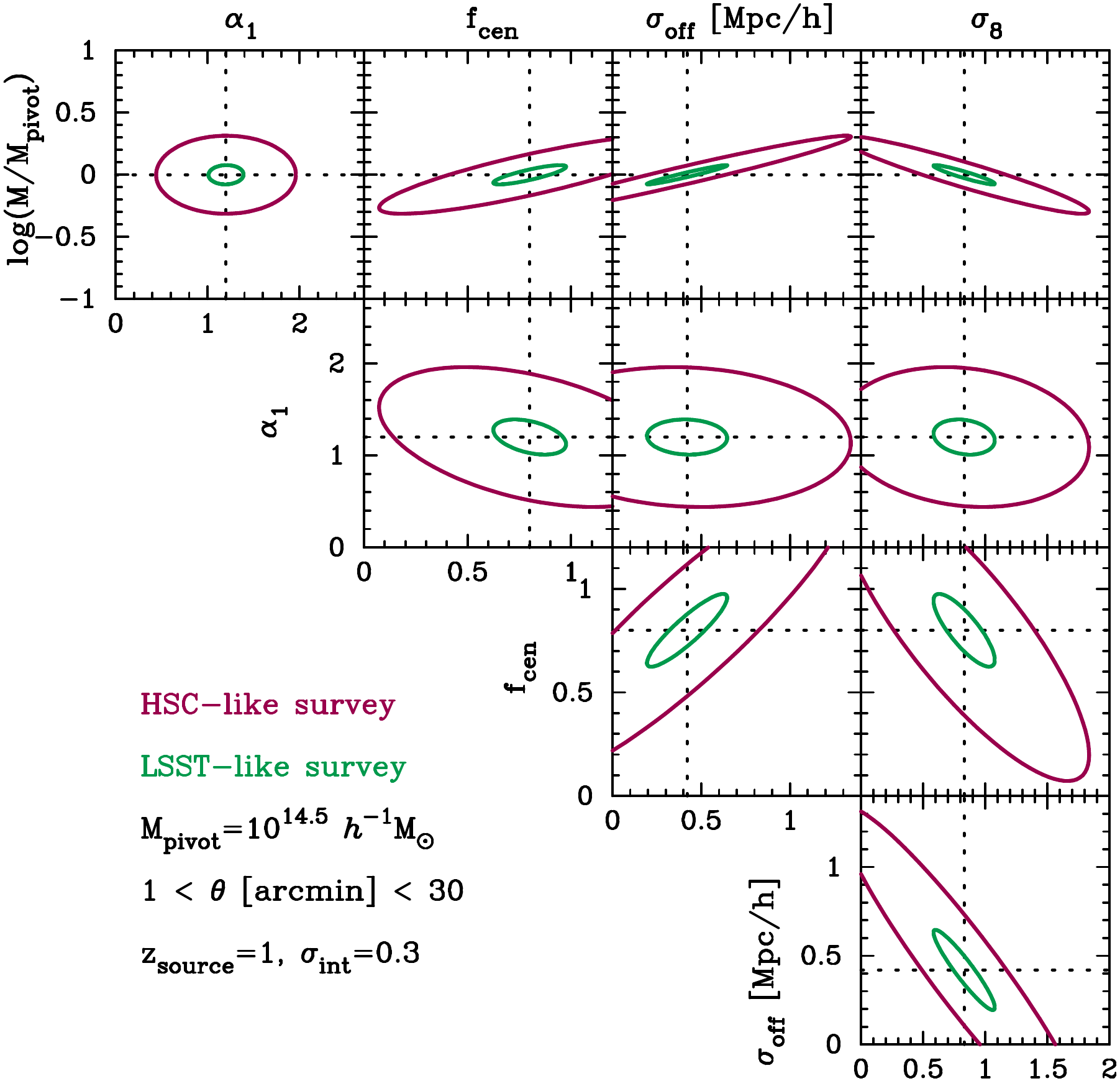}
\caption{
	The expected parameter 
    constraints on stacked 
    lensing signals in ongoing and
    future imaging surveys. 
    Red circles show 
    a 68\% confidence level 
    in parameter spaces of 
    interest for ongoing 
    Subaru HSC-like survey,
    while green ones are 
    for future surveys like LSST.
    For the HSC-like survey, 
    we assume the sky coverage of 
    $1400\, {\rm deg}^{2}$ and 
    the source number density of 
    $30\, {\rm arcmin}^{-2}$.
    For the LSST-like survey,
    we assume 
    the survey area of 
    $20000\, {\rm deg}^{2}$ 
    with the source number density 
    of $50\, {\rm arcmin}^{-2}$.
    In both cases, 
    we consider a cluster sample with 
    mass of $M_{\rm 200m}=10^{14.5}\,h^{-1} M_{\odot}$ 
    at redshift of $0.3$, 
    the source redshift of 1,
    and the intrinsic scatter 
    in galaxy shapes of 0.3.
    }
\label{fig:fisher_circles}
\end{figure*}
%%%%%%%%%%%%%%%%%%%%%%%%%%%%%

%-------------------------------%
\subsection{Modelling stacked weak lensing signal}
\label{subsec:stack_lens_exp}
%-------------------------------%

Stacked lensing analysis of the large-scale structure of the universe is commonly defined by the cross correlation with the shape of background galaxies and the position of foreground objects. This statistical quantity provides a unique method 
of measuring the average mass distribution of foreground objects.
The observable in stacked lensing analysis is 
azimuthally averaged profile of tangential shapes of background galaxies
with respect to the foreground objects denoted as 
$\langle \gamma_{+} \rangle (\theta)$.
Suppose that the intrinsic shape of source galaxies is randomly oriented,
the expected value of $\langle \gamma_{+} \rangle (\theta)$
can be decomposed into three parts:
\beqa
\langle \gamma_{+} \rangle
= f_{\rm cen} \langle \gamma_{+} \rangle_{\rm cen} + (1-f_{\rm cen})\langle \gamma_{+} \rangle_{\rm off} + \langle \gamma_{+} \rangle_{\rm 2halo},
\label{eq:shear_profile}
\eeqa
where the first and second terms on the right hand side of Eq~(\ref{eq:shear_profile})
express the contribution from single dark-matter haloes hosting foreground objects,
while the third term arises from the clustering of neighbouring haloes.
The difference between $\langle \gamma_{+} \rangle_{\rm cen}$
and $\langle \gamma_{+} \rangle_{\rm off}$ comes from 
the imperfect knowledge of centres of host DM haloes.

In optical surveys, the position of the brightest cluster galaxy (BCG) is assumed to be
the centre of the host halo, while some observational studies have shown
that BCGs do not always sit in the centres of haloes 
\citep[e.g.,][]{2010MNRAS.405.2215O, 2012MNRAS.426.2944Z}.
In this paper, we take into account the mis-centering effect of proxy of the halo centre
by introducing \change{the distribution function of the offset between the halo centre and BCG position} as
\beqa
P(R_{\rm off}) = \frac{R_{\rm off}}{\sigma^2_{\rm off}}\exp\left(-\frac{R^{2}_{\rm off}}{2\sigma^2_{\rm off}}\right),
\label{eq:P_miscenter}
\eeqa
where 
$R_{\rm off}$ represents the offset in comoving length scale and $\sigma_{\rm off}$ is the scatter in the probability. 
\change{
In Eq.~(\ref{eq:shear_profile}), the factor $f_{\rm cen}$ represents the fraction of objects located at the true halo centre.}

To predict the first and second terms in Eq~(\ref{eq:shear_profile}),
we assume a smoothly truncated NFW profile as proposed in \citet{2009JCAP...01..015B},
\beqa
\rho_{\rm BMO}(r) = \rho_{\rm NFW}(r)
\left\{\frac{\tau^2_{\rm 200m}}{\left(r/r_{\rm 200m}\right)^2+\tau^2_{\rm 200m}}\right\}^{2}, \label{eq:BMO}
\eeqa
where $r_{\rm 200m}$ is the spherical over-density radius with respect to
$200$ times mean matter density in the Universe and $\tau_{\rm 200m}$ is the characteristic halo truncation radius normalized by $r_{\rm 200m}$, which depends on halo mass and cosmology \citep[e.g.][]{2015MNRAS.449.4264G}.
\citet{2011MNRAS.414.1851O} found that Eq~(\ref{eq:BMO}) can provide 
better agreement with simulated stacked signals compared to simple NFW profile. 
Using these ansatzes, we compute the corresponding lensing signal of $\langle \gamma_{+} \rangle_{\rm cen}$,
while the off-centering term is given by \citep[see][for the derivation]{2011PhRvD..83b3008O} 
\beqa
\langle \gamma_{+} \rangle_{\rm off}(\theta)
= \int \frac{\ell {\rm d}\ell}{2\pi} \, 
\tilde{\kappa}_{\rm BMO}(\ell)\exp\left(-\frac{\sigma^2_{\rm off}\ell^2}{2(1+z_{l})^2 D^2_{l}}\right) J_{2}(\ell \theta),
\eeqa
where 
$\tilde{\kappa}_{\rm BMO}$ is the Fourier component of lensing convergence 
from Eq~(\ref{eq:BMO}),
$z_{l}$ is the redshift of foreground objects,
$D_{l}$ represents the angular diameter distance to foreground objects,
and $J_{2}(x)$ is the second-order Bessel function.

For the third term in Eq~(\ref{eq:shear_profile}), 
we apply the halo-based approach as in \citet{2002PhR...372....1C}
and use the following expression \citep[see also e.g., ][]{2011MNRAS.414.1851O}
\beqa
\langle \gamma_{+} \rangle_{\rm 2halo}(\theta)= 
\frac{b_{h} \Omega_{\rm m0}}{\Sigma_{\rm crit} D^2_{l}}
\int \frac{\ell {\rm d}\ell}{2\pi}\, 
P_{\rm lin}\left(\frac{\ell}{(1+z_{l})D_{l}}, z_{l} \right) J_{2}(\ell \theta),
\eeqa
where $b_{h}$ is the linear halo bias,
$P_{\rm lin}(k, z)$ is the linear matter power spectrum 
for the wavenumber of $k$ at redshift $z$,
and $\Sigma_{\rm crit}$ is the critical mass density for lensing.
In this paper, we adopt the model of $b_{h}$ developed by \citet{2010ApJ...724..878T}.

Hence, the expected signal in stacked lensing analysis contains
the information about the mass density profile on scales smaller than virial radius of halos,
while the cosmological information can be extracted from large-scale signals.
Within our framework,
the relevant parameters in Eq~(\ref{eq:shear_profile}) are: (i) the halo mass ($M_{\rm 200m}$); (ii) the parameters in the mass-profile-MAH relation ($\alpha_{0}$ and $\alpha_{1}$ in Eq~(\ref{eq:general_mass_prof_MAH_relation})); (iii) the parameters for mis-centering probability ($f_{\rm cen}$ and $\sigma_{\rm off}$ in Eq~(\ref{eq:P_miscenter})); (iv) cosmological parameters, especially $\sigma_{8}$.

%-------------------------------%
\subsection{Detectability of baryonic mass contraction in galaxy clusters}
\label{subsec:detect}
%-------------------------------%

Using the theoretical model developed in the previous Section~\ref{subsec:stack_lens_exp}, we forecast the expected observational constraints on the mass-profile-MAH relation with Fisher matrix analysis, which provides an estimate of parameter uncertainties for a given statistical measurement. 

We first introduce the Fisher matrix of stacked lensing signals as 
\beqa
F_{\alpha \beta} &=& 
\frac{1}{2}{\rm Tr}\left[ A_{\alpha}A_{\beta} + C^{-1} H_{\alpha \beta}\right],
\label{eq:Fisher}
\\
A_{\alpha} &=& C^{-1}\partial C/\partial p_{\alpha},
\\
H_{\alpha \beta} &=& 2 (\partial \langle \gamma_{+} \rangle/\partial p_{\alpha})
(\partial \langle \gamma_{+} \rangle/\partial p_{\beta}),
\eeqa
where $\langle \gamma_{+} \rangle$ is the azimuthally averaged profile of tangential shapes of the background lensed galaxies defined in Eq.~(\ref{eq:shear_profile}) that takes into account the off-centering and the two-halo term contributions,
$C$ is the covariance of $\langle \gamma_{+} \rangle$, 
and $p_{\alpha}$ describes the parameter of interest. 
We consider the following parameters:
$\bd{p}=\{\log (M_{\rm 200m} \, [h^{-1}M_{\odot}]), 
\alpha_{1}, f_{\rm cen}, \sigma_{\rm off} \, [h^{-1}{\rm Mpc}], \sigma_{8}\}$.
Here $\alpha_1$ is the slope in the mass-profile-MAH relation in Eq.~(\ref{eq:general_mass_prof_MAH_relation}), $f_{\rm cen}$ and $\sigma_{\rm off}$ characterize the mis-centering of haloes: $f_{\rm cen}$ is the fraction of haloes with centres correctly located in observation (see Section~\ref{subsec:stack_lens_exp}), and $\sigma_{\rm off}$ is the scatter in the mis-centering probability in Eq.~(\ref{eq:P_miscenter}).  
We fix $\alpha_{0}=3$ in Eq~(\ref{eq:general_mass_prof_MAH_relation})
and the lens redshift to be $z_{l}=0.3$.
The marginalized error for each parameter $p_{\alpha}$ 
is given by $\sqrt{F^{-1}_{\alpha \alpha}}$, 
while the un-marginalized error is $1/\sqrt{F_{\alpha \alpha}}$.
We also assume the standard $\Lambda$CDM parameters in
\citet{2016A&A...594A..13P} except for $\sigma_{8}$.
The fiducial parameters of $\bd{p}$ are set to be
$\bd{p}_{\rm fid}
=\{14.5, 1.2, 0.8, 0.42, 0.83\}$.
For the covariance matrix, we adopt the model developed in \citet{2015MNRAS.449.4264G}.
The model includes the statistical uncertainty due to the intrinsic shape, 
the uncorrelated large-scale structure, and mass-dependent term due to intrinsic variations in the mass density profiles. 
When computing the covariance, we perform the linear binning in angular scale $\theta$
from 1 arcmin to 30 arcmin with bin width of 2 arcmin.
To compute the Fisher matrix, 
we only consider the second term in Eq~(\ref{eq:Fisher})
because the mass dependence of the covariance is expected 
to be weak \citep{2015MNRAS.449.4264G}.

\begin{table*}
\centering
\caption{
	A $1\sigma$-level statistical uncertainty
    of our model parameters in hypothetical
    HSC and LSST-like galaxy imaging surveys.
    The number with a round bracket represents
    un-marginalized uncertainty,
    while the marginalized one is for the number
    without bracket.
    The survey parameters (e.g. the sky coverage)
    are the same as 
    in Figure~\ref{fig:fisher_circles}.
    \label{tab:onesigma}
	}
\begin{tabular}{@{}lccccccl}
\hline
\hline
Survey & $\log(M_{\rm 200m}[h^{-1}M_{\odot}])$ & $\alpha_1$ & $f_{\rm cen}$ & $\sigma_{\rm off}\, [h^{-1}{\rm Mpc}]$ & $\sigma_8$ \\ \hline 
HSC-like  & 
0.207 (0.0423) & 0.503 (0.0636) & 0.482 (0.0358) & 0.611 (0.109) & 0.666 (0.0462)  \\ \hline 
LSST-like & 
0.0511 (0.0110) & 0.126 (0.0161) & 0.116 (0.00914) & 0.149 (0.0271) & 0.163 (0.0117)  \\ \hline 
\end{tabular}
\end{table*}

Figure~\ref{fig:fisher_circles} summarizes
the expected constraints of our model parameters
with stacked lensing analysis in ongoing and future imaging surveys.
In this figure, we consider two representative examples, including an ongoing imaging Subaru Hyper-Suprime Cam (HSC) and an upcoming Large Synoptic Survey Telescope (LSST)-like survey. For the HSC-like survey, we assume the sky coverage 
of $1400\, {\rm deg}^{2}$ with the source number density of $30\, {\rm arcmin}^{-2}$.
On the other hand, the LSST-like survey is assumed to cover the sky of $20000\, {\rm deg}^{^2}$
with the source number density of $50\, {\rm arcmin}^{-2}$.
In either case, we set the source redshift to be 1
and the scatter of intrinsic shape in sources of 0.3.

Figure~\ref{fig:fisher_circles} shows the underlying degeneracies among 
our model parameters. 
For instance, more massive haloes tend to show larger signals around the virial radius,
while increasing $\sigma_8$ induces a similar effect on the signal through 
the larger clustering of neighbouring haloes.
Interestingly, the $\alpha_{1}$ parameter 
is found to be almost independent of other parameters, but there exists a degeneracy between $\alpha_1$ and the off-centering effect.
According to our model, 
the $\alpha_1$ parameter can modify the lensing signals at the scale smaller than the virial radius by changing halo concentration, while the off-centering effect can smear the signal at the similar scales. 

After marginalizing over the relevant parameters, we find that the HSC-like survey cannot provide a meaningful constraints on the mass-profile-MAH relation.
Nevertheless, we expect that future wide-area and deep imaging surveys like LSST can measure the mass-profile-MAH relation with good accuracy.
Table~\ref{tab:onesigma} summarizes the expected constraints on the model parameters after marginalization, demonstrating that it is possible to distinguish two models with $\alpha_1=1$ and $\alpha_1=1.2$ at the $1.6\sigma$ level. 
It is worth noting that detectability of baryonic mass contraction in clusters will depend 
on the fiducial value of $\alpha_1$. 
When the fiducial value of $\alpha_1$ is set to be 1.4,
we confirm that LSST-like survey can distinguish the models with $\alpha_1=1$ and $\alpha_1=1.4$ at the $2.8\sigma$ level. 
The constraints as in Table~\ref{tab:onesigma} are found to be insensitive to the choice of 
fiducial $\alpha_1$.
Note that the model with $\alpha_{1}=1$ corresponds to the prediction of dark-matter-only simulations. Hence, the stacked lensing analyses based on future surveys should enable us to measure the baryonic effect on the concentration of galaxy clusters.

%-------------------------------%
\subsection{Bias in parameter estimation}
\label{subsec:bias_param}
%-------------------------------%

Finally, we examine how the parameters in our model will be biased by assuming different baryonic physics, which is encapsulated in the $\alpha_{1}$ parameter. The bias $\delta p_{\alpha}$ for the parameter $p_{\alpha}$ can be expressed as \citep[e.g.,][]{Huterer2006}
\beqa
\delta p_{\alpha} = 
\sum_{\beta}F^{-1}_{\alpha \beta}
\sum_{i,j}
C^{-1}_{ij} \Delta \langle \gamma_{+} \rangle(\theta_{i})
\frac{\partial \langle \gamma_{+} \rangle(\theta_{j})}{\partial p_{\beta}},
\eeqa
where $i$ and $j$ are the indices for binned lensing profiles,
and $\Delta \langle \gamma_{+} \rangle$
describes the difference between the assumed and true models.
In our case, we are interested in how the parameters will be biased when we vary $\alpha_{1}$ from $1.0$, corresponding to no baryonic effects, to $1.2$ the best-fit value from our {\em Omega500} simulations with AGN feedback. Hence we define
\beqa
\Delta \langle \gamma_{+} \rangle (\theta)
\equiv \langle \gamma_{+} \rangle (\theta| \alpha_{1}=1.2)-\langle \gamma_{+} \rangle (\theta| \alpha_{1}=1),
\eeqa
where we use the same parameters as in Section~\ref{subsec:detect} except for $\alpha_{1}$.

Assuming the source redshift to be 1, the scatter of the intrinsic shape in source galaxies is 0.3, and the source number density is $50\, {\rm arcmin}^{-2}$, we find that the bias in $\log M_{\rm 200m}$ is $1.3\times10^{-3}$, and the bias in $\sigma_{8}$ is $3.5\times10^{-3}$.
These biases can be compared to the expected statistical uncertainty given 
by the Fisher matrix.
Assuming the covariance scales with the number of foreground objects,
we find the marginalized statistical uncertainty in $\log M_{\rm 200m}$
is given by $0.85\times \sqrt{10/N_{\rm stack}}$, 
where $N_{\rm stack}$ is the number of foreground objects in the stacked analysis.
Thus, $\log M_{\rm 200m}$ will have a potential bias
on the order of $5\%$ of the statistical uncertainty with $N_{\rm stack}=10,000$.
On the other hand, the marginalized statistical uncertainty in $\sigma_8$ 
is approximated as $2.5\times \sqrt{10/N_{\rm stack}}$,
meaning that the bias in $\sigma_{8}$ is only at the level of $4\%$ 
of the statistical uncertainty for $N_{\rm stack}=10,000$.
Note that the mass of $M_{200m}=10^{14.5}\, h^{-1}M_{\odot}$
approximately corresponds to the typical mass of optically selected galaxy clusters 
and $\sim10,000$ clusters are already available \citep[e.g.,][]{2014ApJ...785..104R}.
Hence, we conclude that the systematic uncertainty due to baryonic effects on stacked lensing analysis will be {\em smaller} than the statistical uncertainty of future galaxy-imaging surveys, such as WFIRST, LSST, and EUCLID. 

%---------------------------------------------------%

\section{CONCLUSIONS}
\label{sec:conclusions}

%---------------------------------------------------%

In this paper, we present a simple model to constrain the effects of baryonic physics in the mass distributions of galaxy clusters, currently one of the major systematic uncertainties in using clusters to constrain cosmology.  
Our model is based on three key assumptions: (1) baryonic processes change the mass distribution via the mass concentration of the halo; (2) there exists a tight `mass profile-MAH' relation, between the mass concentration and the mass accretion history (MAH) for a statistical sample of clusters; and (3) the MAH is determined by mass, redshift and cosmological models, insensitive to baryonic processes. We validated and calibrated our model  using a suite of the {\em Omega500} cosmological hydrodynamical simulations with varying input baryonic physics: gas cooling, star formation and AGN feedback. Our findings are summarized as follows:

\begin{enumerate}

\vspace{2mm}
\item We showed that there exists a fairly tight statistical relation between the enclosed mass density measured within the NFW scaled radius $r_s$, and the critical density of the universe at the time when the main-progenitor mass equals to the mass enclosed within $r_s$, consistent with previous results based on dark-matter-only cosmological simulations. Furthermore, we find that this `mass profile-MAH' relation remains valid in the presence of baryonic physics (cooling, star formation and feedback). Baryonic physics is responsible for changing the slope of the relation (see Figure~\ref{fig:rho_MAH}).

\vspace{2mm}
\item
The changes induced by baryonic physics in the mass profile-MAH relation originate from the increase of the mass concentration, but not from the MAH. In fact, we showed that the MAH is insensitive to baryonic effects (see the coloured points in the right top panel in Figure ~\ref{fig:mass_MAH_comp}). The change in the mass concentration predicted by our model provides a good match to the recent observational constraints from X-ray selected clusters \citep[see Figure~\ref{fig:cprof_Mass_comp_obs_model}]{2016ApJ...821..116U}.

\vspace{2mm}
\item
We showed that cluster mass concentration depends on baryonic physics much more sensitively than on the background cosmology. Precise measurements of the mass concentration with upcoming LSST-like optical survey will allow us to constrain baryonic physics in galaxy clusters (see Figure~\ref{fig:fisher_circles}).

\vspace{2mm}
\item
We applied our model to weak lensing mass measurements and showed that the baryonic effects on mass concentration can introduce the bias in weak-lensing mass estimates at a level of $\sim5\%$ for the realistic marginalized constraint with the stacked lensing measurements for $\sim 10,000$ clusters.
Our model provides a theoretical framework for {\em simultaneously} constraining baryonic physics and cluster masses with upcoming optical surveys. 
\end{enumerate}

This paper presents a theoretical framework for constraining baryonic physics in weak lensing mass, using the mass profile-MAH relation. In particular, the parameters of the relation (i.e., slope and amplitude of the relation) can be used as priors in weak lensing cluster masses. 
\change{The upper limit of the slope can be set to be 1.5, which is expected from our simulation with radiative cooling and star formation (without AGN feedback). 
The lower limit will be 1.0, corresponding to the case without any baryonic effects.}
These two values bracket the uncertainty due to baryonic physics in weak lensing mass estimates. 

Future work should focus on developing and analyzing a larger sample of simulated clusters to verify the model in higher-redshift and lower-mass regimes. In addition, how to relate MAH and other structural properties in mass distribution, e.g., asphericity, still remain uncertain. Addressing these issues is the critical step in understanding the remaining astrophysical uncertainties and hence being able to make accurate and robust interpretations of upcoming cluster surveys.

%%%%%%%%%%%%%%%%%%%%%%%%%%%%%%%%%%%%%%%%%%%%%
% Acknowledgement %%%%%%%%%%%%%%%%%%%%%%%%%%%%%%%%%%
%%%%%%%%%%%%%%%%%%%%%%%%%%%%%%%%%%%%%%%%%%%%%

\section*{acknowledgments} 
We thank Keiichi Umetsu for sharing observational constraints of the mass-concentration relation and useful discussions. We also thank Hironao Miyatake and anonymous referee for helpful comments on the manuscript.
DN and EL acknowledge the support from the NSF grant AST-1412768 and the facilities and staff of the Yale Center for Research Computing.
Numerical computations presented in this paper were in part carried out on the general-purpose PC farm at Center for Computational Astrophysics, CfCA, of National Astronomical Observatory of Japan.

%%%%%%%%%%%%%%%%%%%%%%%%%%%%%%%%%%%%%%%%%%%%%
% REFERENCES %%%%%%%%%%%%%%%%%%%%%%%%%%%%%%%%%%%%
%%%%%%%%%%%%%%%%%%%%%%%%%%%%%%%%%%%%%%%%%%%%%

\bibliographystyle{mnras}
\bibliography{bibtex_library}

\begin{thebibliography}{}
\makeatletter
\relax
\def\mn@urlcharsother{\let\do\@makeother \do\$\do\&\do\#\do\^\do\_\do\%\do\~}
\def\mn@doi{\begingroup\mn@urlcharsother \@ifnextchar [ {\mn@doi@}
  {\mn@doi@[]}}
\def\mn@doi@[#1]#2{\def\@tempa{#1}\ifx\@tempa\@empty \href
  {http://dx.doi.org/#2} {doi:#2}\else \href {http://dx.doi.org/#2} {#1}\fi
  \endgroup}
\def\mn@eprint#1#2{\mn@eprint@#1:#2::\@nil}
\def\mn@eprint@arXiv#1{\href {http://arxiv.org/abs/#1} {{\tt arXiv:#1}}}
\def\mn@eprint@dblp#1{\href {http://dblp.uni-trier.de/rec/bibtex/#1.xml}
  {dblp:#1}}
\def\mn@eprint@#1:#2:#3:#4\@nil{\def\@tempa {#1}\def\@tempb {#2}\def\@tempc
  {#3}\ifx \@tempc \@empty \let \@tempc \@tempb \let \@tempb \@tempa \fi \ifx
  \@tempb \@empty \def\@tempb {arXiv}\fi \@ifundefined
  {mn@eprint@\@tempb}{\@tempb:\@tempc}{\expandafter \expandafter \csname
  mn@eprint@\@tempb\endcsname \expandafter{\@tempc}}}

\bibitem[\protect\citeauthoryear{{Allen}, {Evrard}  \& {Mantz}}{{Allen}
  et~al.}{2011}]{2011ARA&A..49..409A}
{Allen} S.~W.,  {Evrard} A.~E.,   {Mantz} A.~B.,  2011, \mn@doi [ARA\&A]
  {10.1146/annurev-astro-081710-102514}, \href
  {http://adsabs.harvard.edu/abs/2011ARA%26A..49..409A} {49, 409}

\bibitem[\protect\citeauthoryear{{Baltz}, {Marshall}  \& {Oguri}}{{Baltz}
  et~al.}{2009}]{2009JCAP...01..015B}
{Baltz} E.~A.,  {Marshall} P.,   {Oguri} M.,  2009, \mn@doi [JCAP]
  {10.1088/1475-7516/2009/01/015}, \href
  {http://adsabs.harvard.edu/abs/2009JCAP...01..015B} {1, 015}

\bibitem[\protect\citeauthoryear{{Blumenthal}, {Faber}, {Flores}  \&
  {Primack}}{{Blumenthal} et~al.}{1986}]{1986ApJ...301...27B}
{Blumenthal} G.~R.,  {Faber} S.~M.,  {Flores} R.,   {Primack} J.~R.,  1986,
  \mn@doi [ApJ] {10.1086/163867}, \href
  {http://adsabs.harvard.edu/abs/1986ApJ...301...27B} {301, 27}

\bibitem[\protect\citeauthoryear{{Booth} \& {Schaye}}{{Booth} \&
  {Schaye}}{2009}]{booth_etal09}
{Booth} C.~M.,  {Schaye} J.,  2009, \mn@doi [\mnras]
  {10.1111/j.1365-2966.2009.15043.x}, \href
  {http://adsabs.harvard.edu/abs/2009MNRAS.398...53B} {398, 53}

\bibitem[\protect\citeauthoryear{{Broadhurst}, {Umetsu}, {Medezinski}, {Oguri}
  \& {Rephaeli}}{{Broadhurst} et~al.}{2008}]{2008ApJ...685L...9B}
{Broadhurst} T.,  {Umetsu} K.,  {Medezinski} E.,  {Oguri} M.,   {Rephaeli} Y.,
  2008, \mn@doi [ApJL] {10.1086/592400}, \href
  {http://adsabs.harvard.edu/abs/2008ApJ...685L...9B} {685, L9}

\bibitem[\protect\citeauthoryear{{Cooray} \& {Sheth}}{{Cooray} \&
  {Sheth}}{2002}]{2002PhR...372....1C}
{Cooray} A.,  {Sheth} R.,  2002, \mn@doi [Physics Reports]
  {10.1016/S0370-1573(02)00276-4}, \href
  {http://adsabs.harvard.edu/abs/2002PhR...372....1C} {372, 1}

\bibitem[\protect\citeauthoryear{{Duffy}, {Schaye}, {Kay}, {Dalla Vecchia},
  {Battye}  \& {Booth}}{{Duffy} et~al.}{2010}]{2010MNRAS.405.2161D}
{Duffy} A.~R.,  {Schaye} J.,  {Kay} S.~T.,  {Dalla Vecchia} C.,  {Battye}
  R.~A.,   {Booth} C.~M.,  2010, \mn@doi [MNRAS]
  {10.1111/j.1365-2966.2010.16613.x}, \href
  {http://adsabs.harvard.edu/abs/2010MNRAS.405.2161D} {405, 2161}

\bibitem[\protect\citeauthoryear{{Fedeli}}{{Fedeli}}{2012}]{2012MNRAS.424.1244F}
{Fedeli} C.,  2012, \mn@doi [MNRAS] {10.1111/j.1365-2966.2012.21302.x}, \href
  {http://adsabs.harvard.edu/abs/2012MNRAS.424.1244F} {424, 1244}

\bibitem[\protect\citeauthoryear{{Gnedin}, {Kravtsov}, {Klypin}  \&
  {Nagai}}{{Gnedin} et~al.}{2004}]{2004ApJ...616...16G}
{Gnedin} O.~Y.,  {Kravtsov} A.~V.,  {Klypin} A.~A.,   {Nagai} D.,  2004,
  \mn@doi [ApJ] {10.1086/424914}, \href
  {http://adsabs.harvard.edu/abs/2004ApJ...616...16G} {616, 16}

\bibitem[\protect\citeauthoryear{{Gnedin}, {Ceverino}, {Gnedin}, {Klypin},
  {Kravtsov}, {Levine}, {Nagai}  \& {Yepes}}{{Gnedin}
  et~al.}{2011}]{2011arXiv1108.5736G}
{Gnedin} O.~Y.,  {Ceverino} D.,  {Gnedin} N.~Y.,  {Klypin} A.~A.,  {Kravtsov}
  A.~V.,  {Levine} R.,  {Nagai} D.,   {Yepes} G.,  2011, preprint, \href
  {http://adsabs.harvard.edu/abs/2011arXiv1108.5736G} {} (\mn@eprint {arXiv}
  {1108.5736})

\bibitem[\protect\citeauthoryear{{Gruen}, {Seitz}, {Becker}, {Friedrich}  \&
  {Mana}}{{Gruen} et~al.}{2015}]{2015MNRAS.449.4264G}
{Gruen} D.,  {Seitz} S.,  {Becker} M.~R.,  {Friedrich} O.,   {Mana} A.,  2015,
  \mn@doi [MNRAS] {10.1093/mnras/stv532}, \href
  {http://adsabs.harvard.edu/abs/2015MNRAS.449.4264G} {449, 4264}

\bibitem[\protect\citeauthoryear{{Hasselfield} et~al.,}{{Hasselfield}
  et~al.}{2013}]{2013JCAP...07..008H}
{Hasselfield} M.,  et~al., 2013, \mn@doi [\jcap]
  {10.1088/1475-7516/2013/07/008}, \href
  {http://adsabs.harvard.edu/abs/2013JCAP...07..008H} {7, 008}

\bibitem[\protect\citeauthoryear{{Hildebrandt} et~al.,}{{Hildebrandt}
  et~al.}{2017}]{2017MNRAS.465.1454H}
{Hildebrandt} H.,  et~al., 2017, \mn@doi [MNRAS] {10.1093/mnras/stw2805}, \href
  {http://adsabs.harvard.edu/abs/2017MNRAS.465.1454H} {465, 1454}

\bibitem[\protect\citeauthoryear{Hinshaw et~al.,}{Hinshaw
  et~al.}{2013}]{Hinshaw2013}
Hinshaw G.,  et~al., 2013, \mn@doi [ApJS] {10.1088/0067-0049/208/2/19}, 208, 19

\bibitem[\protect\citeauthoryear{Huterer, Takada, Bernstein  \& Jain}{Huterer
  et~al.}{2006}]{Huterer2006}
Huterer D.,  Takada M.,  Bernstein G.,   Jain B.,  2006, \mn@doi [MNRAS]
  {10.1111/j.1365-2966.2005.09782.x}, 366, 101

\bibitem[\protect\citeauthoryear{{Klypin}, {Kravtsov}, {Bullock}  \&
  {Primack}}{{Klypin} et~al.}{2001}]{2001ApJ...554..903K}
{Klypin} A.,  {Kravtsov} A.~V.,  {Bullock} J.~S.,   {Primack} J.~R.,  2001,
  \mn@doi [ApJ] {10.1086/321400}, \href
  {http://adsabs.harvard.edu/abs/2001ApJ...554..903K} {554, 903}

\bibitem[\protect\citeauthoryear{{Komatsu} et~al.,}{{Komatsu}
  et~al.}{2009}]{2009ApJS..180..330K}
{Komatsu} E.,  et~al., 2009, \mn@doi [ApJS] {10.1088/0067-0049/180/2/330},
  \href {http://adsabs.harvard.edu/abs/2009ApJS..180..330K} {180, 330}

\bibitem[\protect\citeauthoryear{{Kravtsov}}{{Kravtsov}}{1999}]{1999PhDT........25K}
{Kravtsov} A.~V.,  1999, PhD thesis, New Mexico State University

\bibitem[\protect\citeauthoryear{{Kravtsov}, {Klypin}  \& {Hoffman}}{{Kravtsov}
  et~al.}{2002}]{2002ApJ...571..563K}
{Kravtsov} A.~V.,  {Klypin} A.,   {Hoffman} Y.,  2002, \mn@doi [ApJ]
  {10.1086/340046}, \href {http://adsabs.harvard.edu/abs/2002ApJ...571..563K}
  {571, 563}

\bibitem[\protect\citeauthoryear{{Ludlow} et~al.,}{{Ludlow}
  et~al.}{2013}]{2013MNRAS.432.1103L}
{Ludlow} A.~D.,  et~al., 2013, \mn@doi [MNRAS] {10.1093/mnras/stt526}, \href
  {http://adsabs.harvard.edu/abs/2013MNRAS.432.1103L} {432, 1103}

\bibitem[\protect\citeauthoryear{{Ludlow}, {Navarro}, {Angulo},
  {Boylan-Kolchin}, {Springel}, {Frenk}  \& {White}}{{Ludlow}
  et~al.}{2014}]{2014MNRAS.441..378L}
{Ludlow} A.~D.,  {Navarro} J.~F.,  {Angulo} R.~E.,  {Boylan-Kolchin} M.,
  {Springel} V.,  {Frenk} C.,   {White} S.~D.~M.,  2014, \mn@doi [MNRAS]
  {10.1093/mnras/stu483}, \href
  {http://adsabs.harvard.edu/abs/2014MNRAS.441..378L} {441, 378}

\bibitem[\protect\citeauthoryear{{Mandelbaum}, {Seljak}  \&
  {Hirata}}{{Mandelbaum} et~al.}{2008}]{2008JCAP...08..006M}
{Mandelbaum} R.,  {Seljak} U.,   {Hirata} C.~M.,  2008, \mn@doi [JCAP]
  {10.1088/1475-7516/2008/08/006}, \href
  {http://adsabs.harvard.edu/abs/2008JCAP...08..006M} {8, 006}

\bibitem[\protect\citeauthoryear{{Mantz}, {Allen}, {Morris}, {Rapetti},
  {Applegate}, {Kelly}, {von der Linden}  \& {Schmidt}}{{Mantz}
  et~al.}{2014}]{2014MNRAS.440.2077M}
{Mantz} A.~B.,  {Allen} S.~W.,  {Morris} R.~G.,  {Rapetti} D.~A.,  {Applegate}
  D.~E.,  {Kelly} P.~L.,  {von der Linden} A.,   {Schmidt} R.~W.,  2014,
  \mn@doi [\mnras] {10.1093/mnras/stu368}, \href
  {http://adsabs.harvard.edu/abs/2014MNRAS.440.2077M} {440, 2077}

\bibitem[\protect\citeauthoryear{{Nagai}, {Kravtsov}  \& {Vikhlinin}}{{Nagai}
  et~al.}{2007}]{2007ApJ...668....1N}
{Nagai} D.,  {Kravtsov} A.~V.,   {Vikhlinin} A.,  2007, \mn@doi [ApJ]
  {10.1086/521328}, \href {http://adsabs.harvard.edu/abs/2007ApJ...668....1N}
  {668, 1}

\bibitem[\protect\citeauthoryear{Navarro, Frenk  \& White}{Navarro
  et~al.}{1996}]{Navarro1996}
Navarro J.~F.,  Frenk C.~S.,   White S. D.~M.,  1996, \mn@doi [ApJ]
  {10.1086/177173}, 462, 563

\bibitem[\protect\citeauthoryear{{Nelson}, {Lau}, {Nagai}, {Rudd}  \&
  {Yu}}{{Nelson} et~al.}{2014}]{2014ApJ...782..107N}
{Nelson} K.,  {Lau} E.~T.,  {Nagai} D.,  {Rudd} D.~H.,   {Yu} L.,  2014,
  \mn@doi [ApJ] {10.1088/0004-637X/782/2/107}, \href
  {http://adsabs.harvard.edu/abs/2014ApJ...782..107N} {782, 107}

\bibitem[\protect\citeauthoryear{{Oguri} \& {Hamana}}{{Oguri} \&
  {Hamana}}{2011}]{2011MNRAS.414.1851O}
{Oguri} M.,  {Hamana} T.,  2011, \mn@doi [MNRAS]
  {10.1111/j.1365-2966.2011.18481.x}, \href
  {http://adsabs.harvard.edu/abs/2011MNRAS.414.1851O} {414, 1851}

\bibitem[\protect\citeauthoryear{{Oguri} \& {Takada}}{{Oguri} \&
  {Takada}}{2011}]{2011PhRvD..83b3008O}
{Oguri} M.,  {Takada} M.,  2011, \mn@doi [Physical Review D]
  {10.1103/PhysRevD.83.023008}, \href
  {http://adsabs.harvard.edu/abs/2011PhRvD..83b3008O} {83, 023008}

\bibitem[\protect\citeauthoryear{{Oguri}, {Takada}, {Okabe}  \&
  {Smith}}{{Oguri} et~al.}{2010}]{2010MNRAS.405.2215O}
{Oguri} M.,  {Takada} M.,  {Okabe} N.,   {Smith} G.~P.,  2010, \mn@doi [MNRAS]
  {10.1111/j.1365-2966.2010.16622.x}, \href
  {http://adsabs.harvard.edu/abs/2010MNRAS.405.2215O} {405, 2215}

\bibitem[\protect\citeauthoryear{{Oguri}, {Bayliss}, {Dahle}, {Sharon},
  {Gladders}, {Natarajan}, {Hennawi}  \& {Koester}}{{Oguri}
  et~al.}{2012}]{2012MNRAS.420.3213O}
{Oguri} M.,  {Bayliss} M.~B.,  {Dahle} H.,  {Sharon} K.,  {Gladders} M.~D.,
  {Natarajan} P.,  {Hennawi} J.~F.,   {Koester} B.~P.,  2012, \mn@doi [MNRAS]
  {10.1111/j.1365-2966.2011.20248.x}, \href
  {http://adsabs.harvard.edu/abs/2012MNRAS.420.3213O} {420, 3213}

\bibitem[\protect\citeauthoryear{{Okabe}, {Takada}, {Umetsu}, {Futamase}  \&
  {Smith}}{{Okabe} et~al.}{2010}]{2010PASJ...62..811O}
{Okabe} N.,  {Takada} M.,  {Umetsu} K.,  {Futamase} T.,   {Smith} G.~P.,  2010,
  \mn@doi [PASJ] {10.1093/pasj/62.3.811}, \href
  {http://adsabs.harvard.edu/abs/2010PASJ...62..811O} {62, 811}

\bibitem[\protect\citeauthoryear{{Planck Collaboration} et~al.,}{{Planck
  Collaboration} et~al.}{2016a}]{2016A&A...594A..13P}
{Planck Collaboration} et~al., 2016a, \mn@doi [A\&A]
  {10.1051/0004-6361/201525830}, \href
  {http://adsabs.harvard.edu/abs/2016A%26A...594A..13P} {594, A13}

\bibitem[\protect\citeauthoryear{{Planck Collaboration} et~al.,}{{Planck
  Collaboration} et~al.}{2016b}]{2016A&A...594A..24P}
{Planck Collaboration} et~al., 2016b, \mn@doi [A\&A]
  {10.1051/0004-6361/201525833}, \href
  {http://adsabs.harvard.edu/abs/2016A%26A...594A..24P} {594, A24}

\bibitem[\protect\citeauthoryear{{Press}, {Teukolsky}, {Vetterling}  \&
  {Flannery}}{{Press} et~al.}{1992}]{1992nrfa.book.....P}
{Press} W.~H.,  {Teukolsky} S.~A.,  {Vetterling} W.~T.,   {Flannery} B.~P.,
  1992, {Numerical Recipes in FORTRAN. The Art of Scientific Computing}.
Cambridge: University Press, |c1992, 2nd ed.

\bibitem[\protect\citeauthoryear{{Rudd}, {Zentner}  \& {Kravtsov}}{{Rudd}
  et~al.}{2008}]{2008ApJ...672...19R}
{Rudd} D.~H.,  {Zentner} A.~R.,   {Kravtsov} A.~V.,  2008, \mn@doi [ApJ]
  {10.1086/523836}, \href {http://adsabs.harvard.edu/abs/2008ApJ...672...19R}
  {672, 19}

\bibitem[\protect\citeauthoryear{{Ryden} \& {Gunn}}{{Ryden} \&
  {Gunn}}{1987}]{1987ApJ...318...15R}
{Ryden} B.~S.,  {Gunn} J.~E.,  1987, \mn@doi [ApJ] {10.1086/165349}, \href
  {http://adsabs.harvard.edu/abs/1987ApJ...318...15R} {318, 15}

\bibitem[\protect\citeauthoryear{{Rykoff} et~al.,}{{Rykoff}
  et~al.}{2014}]{2014ApJ...785..104R}
{Rykoff} E.~S.,  et~al., 2014, \mn@doi [ApJ] {10.1088/0004-637X/785/2/104},
  \href {http://adsabs.harvard.edu/abs/2014ApJ...785..104R} {785, 104}

\bibitem[\protect\citeauthoryear{{Schaller} et~al.,}{{Schaller}
  et~al.}{2015a}]{2015MNRAS.451.1247S}
{Schaller} M.,  et~al., 2015a, \mn@doi [MNRAS] {10.1093/mnras/stv1067}, \href
  {http://adsabs.harvard.edu/abs/2015MNRAS.451.1247S} {451, 1247}

\bibitem[\protect\citeauthoryear{{Schaller} et~al.,}{{Schaller}
  et~al.}{2015b}]{2015MNRAS.452..343S}
{Schaller} M.,  et~al., 2015b, \mn@doi [MNRAS] {10.1093/mnras/stv1341}, \href
  {http://adsabs.harvard.edu/abs/2015MNRAS.452..343S} {452, 343}

\bibitem[\protect\citeauthoryear{{Tinker}, {Robertson}, {Kravtsov}, {Klypin},
  {Warren}, {Yepes}  \& {Gottl{\"o}ber}}{{Tinker}
  et~al.}{2010}]{2010ApJ...724..878T}
{Tinker} J.~L.,  {Robertson} B.~E.,  {Kravtsov} A.~V.,  {Klypin} A.,  {Warren}
  M.~S.,  {Yepes} G.,   {Gottl{\"o}ber} S.,  2010, \mn@doi [ApJ]
  {10.1088/0004-637X/724/2/878}, \href
  {http://adsabs.harvard.edu/abs/2010ApJ...724..878T} {724, 878}

\bibitem[\protect\citeauthoryear{{Umetsu} et~al.,}{{Umetsu}
  et~al.}{2014}]{2014ApJ...795..163U}
{Umetsu} K.,  et~al., 2014, \mn@doi [\apj] {10.1088/0004-637X/795/2/163}, \href
  {http://adsabs.harvard.edu/abs/2014ApJ...795..163U} {795, 163}

\bibitem[\protect\citeauthoryear{{Umetsu}, {Zitrin}, {Gruen}, {Merten},
  {Donahue}  \& {Postman}}{{Umetsu} et~al.}{2016}]{2016ApJ...821..116U}
{Umetsu} K.,  {Zitrin} A.,  {Gruen} D.,  {Merten} J.,  {Donahue} M.,
  {Postman} M.,  2016, \mn@doi [ApJ] {10.3847/0004-637X/821/2/116}, \href
  {http://adsabs.harvard.edu/abs/2016ApJ...821..116U} {821, 116}

\bibitem[\protect\citeauthoryear{{Velliscig}, {van Daalen}, {Schaye},
  {McCarthy}, {Cacciato}, {Le Brun}  \& {Dalla Vecchia}}{{Velliscig}
  et~al.}{2014}]{2014MNRAS.442.2641V}
{Velliscig} M.,  {van Daalen} M.~P.,  {Schaye} J.,  {McCarthy} I.~G.,
  {Cacciato} M.,  {Le Brun} A.~M.~C.,   {Dalla Vecchia} C.,  2014, \mn@doi
  [MNRAS] {10.1093/mnras/stu1044}, \href
  {http://adsabs.harvard.edu/abs/2014MNRAS.442.2641V} {442, 2641}

\bibitem[\protect\citeauthoryear{{Vikhlinin} et~al.,}{{Vikhlinin}
  et~al.}{2009}]{2009ApJ...692.1060V}
{Vikhlinin} A.,  et~al., 2009, \mn@doi [\apj] {10.1088/0004-637X/692/2/1060},
  \href {http://adsabs.harvard.edu/abs/2009ApJ...692.1060V} {692, 1060}

\bibitem[\protect\citeauthoryear{{Wechsler}, {Bullock}, {Primack}, {Kravtsov}
  \& {Dekel}}{{Wechsler} et~al.}{2002}]{2002ApJ...568...52W}
{Wechsler} R.~H.,  {Bullock} J.~S.,  {Primack} J.~R.,  {Kravtsov} A.~V.,
  {Dekel} A.,  2002, \mn@doi [ApJ] {10.1086/338765}, \href
  {http://adsabs.harvard.edu/abs/2002ApJ...568...52W} {568, 52}

\bibitem[\protect\citeauthoryear{{Zitrin}, {Bartelmann}, {Umetsu}, {Oguri}  \&
  {Broadhurst}}{{Zitrin} et~al.}{2012}]{2012MNRAS.426.2944Z}
{Zitrin} A.,  {Bartelmann} M.,  {Umetsu} K.,  {Oguri} M.,   {Broadhurst} T.,
  2012, \mn@doi [MNRAS] {10.1111/j.1365-2966.2012.21886.x}, \href
  {http://adsabs.harvard.edu/abs/2012MNRAS.426.2944Z} {426, 2944}

\bibitem[\protect\citeauthoryear{{de Haan} et~al.,}{{de Haan}
  et~al.}{2016}]{2016ApJ...832...95D}
{de Haan} T.,  et~al., 2016, \mn@doi [\apj] {10.3847/0004-637X/832/1/95}, \href
  {http://adsabs.harvard.edu/abs/2016ApJ...832...95D} {832, 95}

\bibitem[\protect\citeauthoryear{{van den Bosch}}{{van den
  Bosch}}{2002}]{2002MNRAS.331...98V}
{van den Bosch} F.~C.,  2002, \mn@doi [MNRAS]
  {10.1046/j.1365-8711.2002.05171.x}, \href
  {http://adsabs.harvard.edu/abs/2002MNRAS.331...98V} {331, 98}

\makeatother
\end{thebibliography}

%%%%%%%%%%%%%%%%%%%%%%%%%%%%%
%\clearpage
\appendix

%-------------------------------%
\section{Model of Mass Concentration and Accretion History}
\label{app:mah_model}
%-------------------------------%

\change{
Here we summarize how to compute the mass concentration $c$ for arbitrary masses, redshifts,
and cosmologies once Eq.~(\ref{eq:cs_relation}) is determined.
}

For a theoretical model of $c_{\rm MAH}$,
We adopt a simple model of MAH proposed by \citet{2002MNRAS.331...98V}. 
In this model, the MAH of DM  halos can be expressed as
\beqa
\log\left(\frac{M_{\rm prog}(z)}{M_{0}}\right)
=-0.301\left[\frac{\log(1+z)}{\log(1+z_{f})}\right]^{\chi}, \label{eq:model_MAH}
\eeqa
where 
$M_{0} = M_{\rm prog}(z=0)$,
$z_{f}$ is a characteristic formation redshift,
and $\chi$ is a parameter that depends on cosmology and mass.
These two parameters are defined as
\beqa
\chi &=& 1.211+1.858\log(1+z_{f})+0.308\Omega^2_{\Lambda} \nonumber \\
&& -0.032\log(M_{0}/[10^{11}\, h^{-1}M_{\odot}]),
\eeqa  
\beqa 
\delta_{\rm sph}/D(z_f) = \delta_{\rm sph} + 0.477 \sqrt{2\left[\sigma^2(fM_0)-\sigma^2(M_0)\right]},
\eeqa
where 
$\delta_{\rm sph}=1.686$, 
$D(z)$ is the linear growth factor and $\sigma^2(M)$ is 
the mass variance in spheres of mass $M$ 
computed from the linear power spectrum at $z=0$.
The parameter $f$ depends on the spherical over-density 
definition of $M_{\rm prog}$ and we find $f=0.1$ provides 
a reasonable fit to our simulation results for $\Delta=500$.
Since Eq.~(\ref{eq:model_MAH}) can be fitted to the NFW form
\citep[see also][]{2014MNRAS.441..378L}, we derive the corresponding 
$c_{\rm MAH}$ at $z=0$ for the MAH given by Eq.~(\ref{eq:model_MAH}).

For non-zero redshift ($z_{o}\neq0$),
we will derive the concentration $c_{\rm MAH}$ of DM halos 
by assuming the following MAH in the range of $z>z_{o}$:
\beqa
\log\left(\frac{M_{\rm prog}(z)}{M_{\rm prog}(z_{o})}\right)
&=& -0.301 
\Bigg\{
\left[ \frac{\log(1+z)}{\log(1+z_f)}\right]^{\chi} \nonumber \\
&&
\,\,\,\,\,
\,\,\,\,\,
-\left[ \frac{\log(1+z_o)}{\log(1+z_f)}\right]^{\chi}
\Bigg\}. \label{eq:model_MAH_diff_z}
\eeqa
Since Eq.~(\ref{eq:model_MAH_diff_z}) 
can also be expressed as a function of 
$\rho_{\rm crit}(z)/\rho_{\rm crit}(z_o)$,
we can derive $c_{\rm MAH}$ by using the NFW profile in Eq.~(\ref{eq:MAH_scaled}).
Using the relation between $c_{\rm MAH}$ and $c$ given in 
Eq.~(\ref{eq:cs_relation}),
we evaluate $c$ at the redshift of $z_o$ for a given $M_{\rm prog}(z=0)$.
Once determining $c$ at $z=z_{o}$, 
we can then derive the spherical over-density mass $M_{\Delta}(z_o)$ by solving 
\beqa
\frac{M_{\rm prog}(z_o)}{M_{\Delta}(z_o)} 
&=& \frac{Y(cx)}{Y(c)}, \label{eq:mdelta_z1} \\
\frac{M_{\rm prog}(z_o)}{M_{\rm prog}(z=0)}
&=&-0.301\left[\frac{\log(1+z_o)}{\log(1+z_{f})}\right]^{\chi}, 
\label{eq:mdelta_z2} \\
M_{\rm prog}(z_o)
&=& \frac{4\pi}{3} \Delta^{\prime} \, \rho_{\rm crit}(z_o) r^3_{\Delta^{\prime}}(z_o), \label{eq:mdelta_z3}
\eeqa
where $x=r_{\Delta^{\prime}}(z_o)/r_{\Delta}(z_o)$. 
\change{
Note that the parameters in Eq.~(\ref{eq:cs_relation}) depend on the spherical over-density definition 
of $M_{\rm prog}$. Furthermore, it is possible to define $M_{\rm prog}$ using different over-density parameters in $M_{\Delta}$ of interest in general. Hence, Eq~(\ref{eq:mdelta_z3}) is needed to close the set of equations.}

\vspace{2mm}
%We now have a procedure for computing $M_{\Delta c}$ for a given set of 
%$M_{\rm prog}(z=0)$, $z_o$, and $c$.
%The procedure to derive $c$ consists of four steps: 
\change{
We now have a four-step procedure to derive $c$ as follows: (i) modeling MAH of a DM halo for a given present-day mass $M_{\rm prog}(z=0)$, redshift $z_o$, and cosmology; (ii) determining $c_{\rm MAH}$ for the given MAH $M_{\rm prog}(z)$ using Eq.~(\ref{eq:MAH_scaled}); (iii) translating $c_{\rm MAH}$ to the concentration of mass density profile $c$ using Eq.~(\ref{eq:cs_relation}); (iv) deriving the spherical mass $M_{\Delta}$ at redshift of $z_{o}$ by solving Eqs~(\ref{eq:mdelta_z1}), (\ref{eq:mdelta_z2}), and (\ref{eq:mdelta_z3}).}

%\begin{enumerate}
%\vspace{2mm}
%\change{
%\item[(i)] model MAH of DM halo for a given present-day mass $M_{\rm prog}(z=0)$, redshift $z_o$, and cosmology as shown above
%}

%\vspace{2mm}
%\change{
%\item[(ii)] find the NFW expression of modelled MAH $M_{\rm prog}(z)$ and determine $c_{\rm MAH}$ with Eq.~(\ref{eq:MAH_scaled}) 
%}

%\vspace{2mm}
%\change{
%\item[(iii)] translate $c_{\rm MAH}$ to the concentration of mass density profile $c$ as in Eq.~(\ref{eq:cs_relation}) 
%}

%\vspace{2mm}
%\change{
%\item[(iv)] derive the spherical mass $M_{\Delta}$ at redshift of $z_{o}$ by solving Eqs~(\ref{eq:mdelta_z1}), (\ref{eq:mdelta_z2}), and (\ref{eq:mdelta_z3}).}
%\end{enumerate}
 
\section{Comparisons of simulated mass profile and contraction model}
\label{app:mass_contraction_model}

%%%%%%%%%%%%%%%%%%%%%%%%%%%%%
\begin{figure}
\centering
\includegraphics[width=0.8\columnwidth, bb=0 0 511 538]
{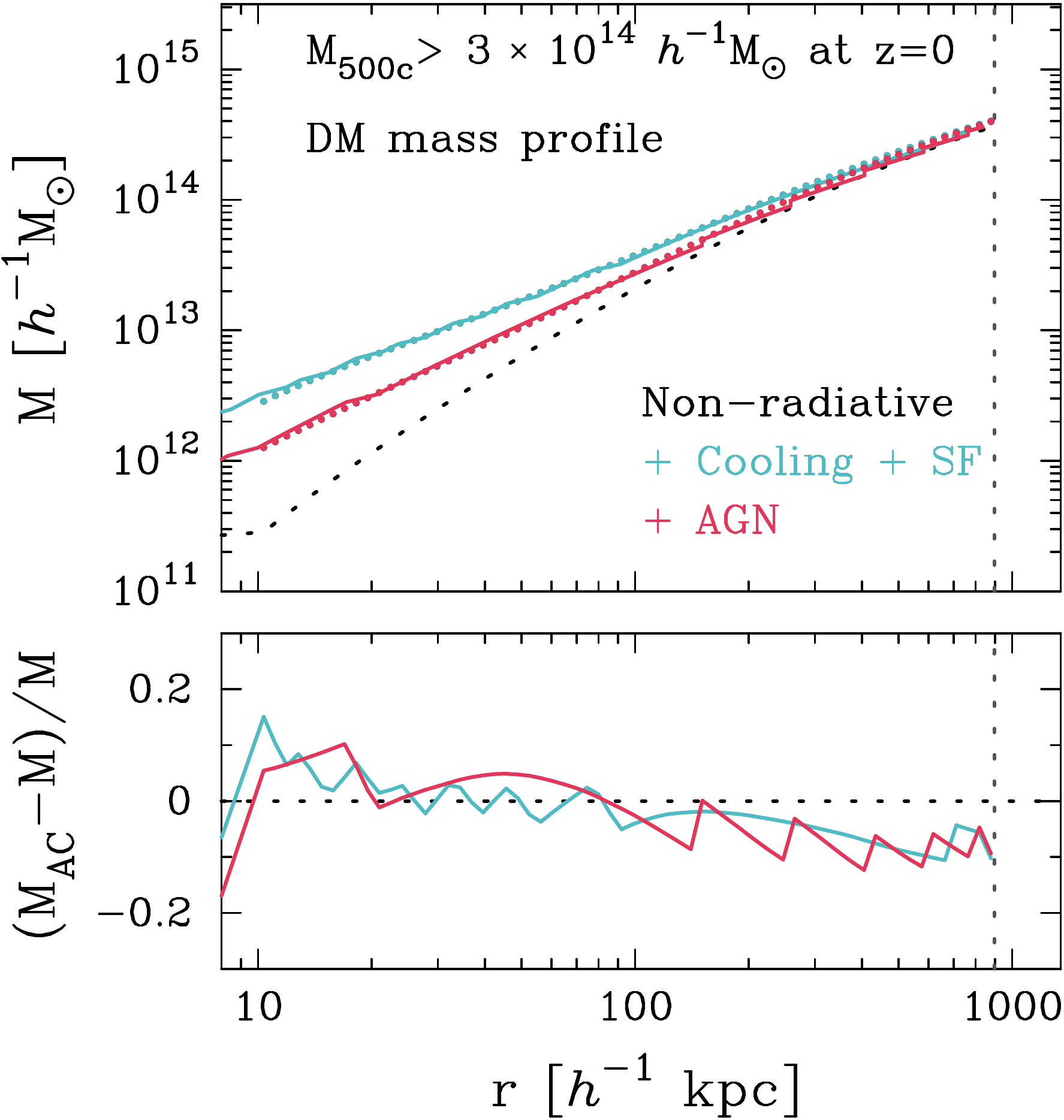}
\caption{
	Comparison of enclosed dark-matter profile
	in our simulations and modified contraction model.
	The dashed line represents the DM profile in the NR run, 
	while the cyan and red points are for the CSF and AGN runs, respectively.
	Colored lines show the predictions by modified contraction model.
	\change{Using the least chi-squared fitting, we determined the best-fit parameters of the Eq.~(\ref{eq:rbar}) to be ${\cal A}=1.12$ and $w=0.83$ for the CSF run and ${\cal A}=1.11$ and $w=0.78$ for the AGN run.}
	As shown in the bottom panel, 
	the model can explain the average dark-matter mass profile for mass-limited sample
	in our simulations with a level of 0.1 dex.
	Note that the horizontal dashed line in each panel
	represents the spherical over-density radius of $r_{500c}$
	for $M_{\rm 500c}=3\times10^{14}\, h^{-1}M_{\odot}$.
	}
\label{fig:halo_contraction}
\end{figure}

Here we present the comparison of mass profile in our simulations
with modified contraction model by \citet{2011arXiv1108.5736G}.
%Suppose that the shell enclosing mass shell at initial location $r_{\rm i}$
%and the final location $r_{\rm f}$ after contraction,
\change{
Suppose that the mass profile consists of spherical shells which contract in radius but do not cross each other, where the initial location of the mass shell is $r_{\rm i}$ and the final location after contraction is $r_{\rm f}$.}
\citet{2004ApJ...616...16G} have proposed the following expression to reproduce the simulated contraction due to \change{baryon dissipation};
\beqa
\left[M_{\rm DM}(\bar{r}_{\rm i})
+M_{\rm gas, i}(\bar{r}_{\rm i})\right] \, r_{\rm i}
= \left[M_{\rm DM}(\bar{r}_{\rm i})
+M_{\rm gas, f}(\bar{r}_{\rm f})\right] \, r_{\rm f}, \label{eq:MAC}
\eeqa
where $M_{\rm DM}(r)$ is the enclosed dark-matter mass within the radius of $r$,
$M_{\rm gas, i}(r)$ is the total initial baryon mass within $r$,
and $M_{\rm gas, f}(r)$ is the total final baryon mass within $r$.
In Eq.~(\ref{eq:MAC}), $\bar{r}$ represents
the orbit-averaged radius for particles currently located at radius $r$
and it can be expressed as
\beqa
\frac{\bar{r}}{r_{\rm 500c}} = {\cal A}\left(\frac{r}{r_{\rm 500c}}\right)^{w},
\label{eq:rbar}
\eeqa
where the case of ${\cal A}=w=1$ corresponds to simple adiabatic contraction model
\citep{1986ApJ...301...27B, 1987ApJ...318...15R}.
To compare our simulated mass profile and Eq.~(\ref{eq:MAC}),
we work with a mass-limited sample with 
$M_{\rm 500c}>3\times10^{14}\, h^{-1}M_{\odot}$ at $z=0$ for three different runs.
We compute the average dark-matter and baryon mass profiles for each run.
We then compute the final location $r_{\rm f}$ in the contraction model
assuming 
$M_{\rm DM}(r)$ is set to be the average dark-matter profile in the NR run, 
$M_{\rm gas, i}$ is the average baryon mass profile in the NR run, 
and $M_{\rm gas, f}$ is the average baryon mass profile in the CSF or AGN runs.
The predicted mass profile by Eq.~(\ref{eq:MAC}) is given by
$M_{\rm DM, AC}(r) = M_{\rm DM}(r_{\rm f})$, and this will be compared
with the simulated dark-matter mass profile in CSF or AGN runs.

Figure~\ref{fig:halo_contraction} shows the comparison of dark-matter mass profile in our simulations with the predicted one. The cyan and red points are for simulated mass profile, while the colored lines are predicted mass $M_{\rm DM, AC}$. We found that the modified contraction model as in Eq.~(\ref{eq:MAC})
can explain the simulated dark-matter enclosed mass with the level of $\simeq0.1$ dex in the range of $10^{-2}<r/r_{\rm 500c}<1$. The model can capture the feature of contraction in mass profile for a given baryonic mass profile.

\end{document}